\documentclass[twocolumn]{aastex62}
\usepackage{bm}
\usepackage{blindtext}
\usepackage{hyperref}
\newcommand{\bjdtdb}{${\rm {BJD_{TDB}}}$}

\newcommand{\rsun}{${\rm R}_\Sun$}

\newcommand{\mj}{${\,{\rm M}_{\rm J}}$}
\newcommand{\rj}{${\,{\rm R}_{\rm J}}$}

\newcommand{\three}{3.6\,$\mu$m\ }
\newcommand{\four}{4.5\,$\mu$m\ }
\newcommand{\threealt}{3.6\,$\mu$m}
\newcommand{\fouralt}{4.5\,$\mu$m}
\newcommand{\um}{$\mu$m}

\newcommand{\methane}{$\mathrm{CH}_4$}

\begin{document}

\title{Spitzer Phase Curves of KELT-1b and the Signatures of Nightside Clouds in Thermal Phase Observations}

\author{Thomas G.\ Beatty}
\affiliation{Department of Astronomy and Steward Observatory, University of Arizona, Tucson, AZ 85721; tgbeatty@email.arizona.edu}

\author{Mark S. Marley}
\affiliation{NASA Ames Research Center, Moffett Field, CA 94035}

\author{B. Scott Gaudi}
\affiliation{Department of Astronomy, The Ohio State University, Columbus, OH 43210}

\author{Knicole D. Col\'{o}n}
\affiliation{NASA Goddard Space Flight Center, Exoplanets and Stellar Astrophysics Laboratory (Code 667), Greenbelt, MD 20771}

\author{Jonathan J. Fortney}
\affiliation{Department of Astronomy \& Astrophysics, University of California, Santa Cruz, CA 95064}

\author{Adam P. Showman}
\affiliation{Department of Planetary Sciences and Lunar and Planetary Laboratory, University of Arizona, Tucson, AZ 85721}

\shorttitle{Spitzer Phase Curves of KELT-1b}
\shortauthors{Beatty et al.}


\begin{abstract}
We observed two full orbital phase curves of the transiting brown dwarf KELT-1b, at \three and \fouralt, using the Spitzer Space Telescope. Combined with previous eclipse data from \cite{beatty2014}, we strongly detect KELT-1b's phase variation as a single sinusoid in both bands, with amplitudes of $964\pm36$\,ppm at \three and $979\pm54$\,ppm at \fouralt, and confirm the secondary eclipse depths measured by \cite{beatty2014}. We also measure noticeable Eastward hotspot offsets of $28.4\pm3.5$ degrees at \three and $18.6\pm5.2$ degrees at \fouralt. Both the day-night temperature contrasts and the hotspot offsets we measure are in line with the trends seen in hot Jupiters \citep[e.g.][]{crossfield2015}, though we disagree with the recent suggestion of an offset trend by \cite{zhang2018}. Using an ensemble analysis of Spitzer phase curves, we argue that nightside clouds are playing a noticeable role in modulating the thermal emission from these objects, based on: 1) the lack of a clear trend in phase offsets with equilibrium temperature, 2) the sharp day-night transitions required to have non-negative intensity maps, which also resolves the inversion issues raised by \cite{keating2017}, 3) the fact that all the nightsides of these objects appear to be at roughly the same temperature of 1000\,K, while the dayside temperatures increase linearly with equilibrium temperature, and 4) the trajectories of these objects on a Spitzer color-magnitude diagram, which suggest colors only explainable via nightside clouds.
\end{abstract}

\section{Introduction}

Orbital phase curve observations are one of the few ways in which we can directly investigate the global climates of exoplanets. This is particularly important for strongly irradiated planets such as hot Jupiters, since there can be temperature differences of over one thousand degrees between their day- and nightsides. This is believed to drive noticeable atmospheric composition changes between the two hemispheres, to say nothing of radically altering the vertical temperature structure as a function of longitude. 

One critically important component of the day-to-night changes in hot Jupiters is the possible formation and clearing of clouds on their night- and daysides. Though the possible role of clouds in exoplanet atmospheres has been appreciated for quite some time \citep{burrows1997,marley1999}, much of the 3D modeling of hot Jupiter atmospheres has assumed they are cloud free \citep[e.g.,][]{showman2008,kataria2016}. As a practical matter, this is due to the Herculean task of constructing accurate 3D global circulation models (GCMs) that properly deal with ``just'' dynamics and radiative transfer \citep{showman2008}. The modeling effort to add self-consistent cloud physics, including condensation processes and size distributions, that link to the established radiative and dynamics codes is just getting underway \citep[e.g.][]{lee2016}.

As a result, the results of Spitzer phase curve observations are usually contextualized using a framework of competing ``thermal-only'' effects within hot Jupiters' atmospheres. For example, there is now a well-established trend that hot Jupiters with higher zero-albedo complete heat redistribution equilibrium temperatures (i.e., higher stellar irradiation) also show a higher temperature contrast between their day- and nightsides \citep{perezbecker2013}. Both the early theoretical work that predicted this trend \citep{showman2002} and more recent GCM analyses \citep{komacek2016} explain this using differences in the atmospheric radiative timescales and the atmospheric advective \citep{showman2002} or drag \citep{komacek2016} timescales. Put another way, the temperature difference between the day- and nightside of a hot Jupiter is determined by the balance of how fast the atmosphere cools and how fast it moves heat to the nightside.

In all these analyses it has been made clear that the inclusion of clouds has the potential to strongly affect the results of the simulations, and recently more effort has been devoted to incorporating clouds into GCMs. In particular, \cite{parmentier2013}, \cite{lee2016}, and \cite{macdonald2017} have all found that 3D or 2D atmospheric models of HD 209458b that include cloud physics do a better job of replicating that planet's emission spectrum than cloud free models. Recently, \cite{powell2018} described a general atmosphere model that couples dynamics, radiative transfer, and cloud physics -- and which predicts that hot Jupiters should generally possess a nightside cloud deck.

Observationally, the presence of high altitude clouds along planetary terminators has been evident in transmission spectroscopy measurements for some time \citep[e.g.][]{kreidberg2014}, but the signatures of clouds in emission measurements have been more difficult to see. This is because the daysides of hot Jupiters -- which provide us with our best emission spectra -- are believed to be mostly cloud-free \citep{parmentier2016}, though \cite{beatty2017a} recently inferred nightside TiO condensation on Kepler-13Ab based on that planet's dayside emission.

The consideration of global, day and night, cloud coverage has largely been driven by the availability of red-optical phase curve data from Kepler. Initially, the phases curves of some individual planets showed clear Westward hotspot offsets that seem to strongly indicate clouds \citep[e.g.][]{demory2013}. \cite{parmentier2016} performed an ensemble analysis of Kepler phase curve results and found that clouds were generally required to explain not only the reflection signals themselves, but also how the amplitude and offsets of the planetary phase curves changed with temperature in the Kepler bandpass.

One notable result from the Kepler phase curve observations is the apparently variable cloud cover on HAT-P-7b \citep{armstrong2016}. Four years of Kepler data show that not only does the location of maximum flux shift by up to 80 degrees over hundreds of days, but the shape of the phase curve itself is also variable on the same time scale. \cite{armstrong2016} interpreted this as changes in the weather on HAT-P-7b as the planetary cloud cover changed both its extent and its location. Interestingly, their toy model to explain the observations required that the observable thermal emission from HAT-P-7b was being altered by the variable clouds -- and not just the reflected light signal.

In the infrared, \cite{mendonca2018} reanalyzed \cite{stevenson2017}'s \three and \four Spitzer phase curves of WASP-43b using a toy-model for clouds, by approximating their presence as a constant additional atmospheric opacity on the planetary nightside. \cite{mendonca2018} found that their cloudy simulations agreed more closely with the low observed nightside thermal emission, as the inclusion of clouds caused the modeled nightside flux to be significantly lower than predicted in cloud-free atmosphere.

The \cite{armstrong2016} and \cite{mendonca2018} results indicate that even at the thermal infrared wavelengths probed by Spitzer, we should be considering how clouds modulate the thermal emission of hot Jupiters. This was recently, again, evident in \cite{dang2018}'s single \four phase curve of CoRoT-2b, which displayed a Westward hotspot offset, and was taken to be evidence for clouds affecting the dayside thermal emission of CoRoT-2b. Additionally, if the long-term cloud variation that \cite{armstrong2016} saw in HAT-P-7b is representative of the entire population of hot Jupiters, then there is the distinct possibility that Spitzer phase curve results are observing the combination of short-term weather effects on top of the equilibrium climates of hot Jupiters.

\begin{figure*}[t]
\vskip 0.00in
\includegraphics[width=1.05\linewidth,clip]{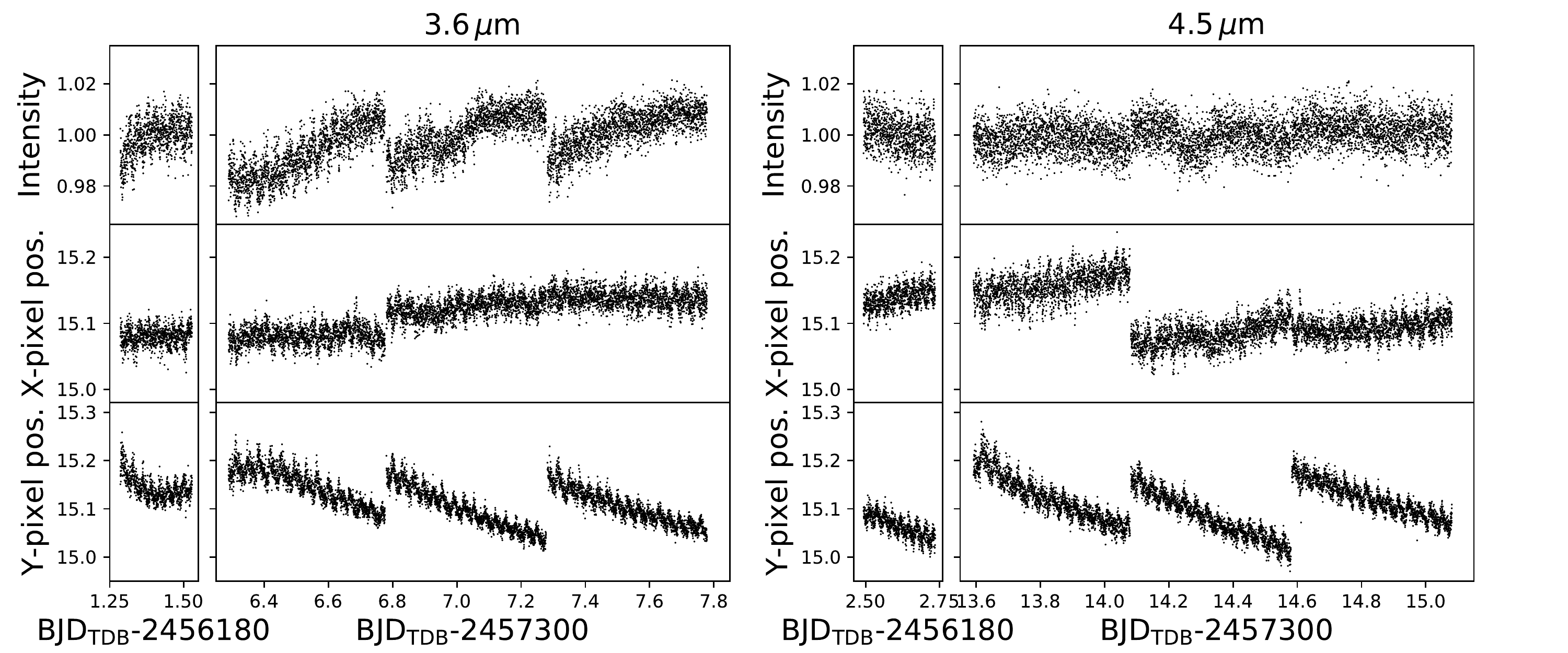}
\vskip -0.0in
\caption{The raw photometry we used for our analysis combined new observations covering an entire orbit at \three and \four with \three and \four eclipse photometry previously analyzed in \cite{beatty2014}. Both data sets display correlations between the measured intensity and the x- and y-pixel position of the stellar centroid, which are typical features of \three and \four Spitzer photometry.}
\label{rawplot}
\end{figure*}

To investigate the role of thermal-only effects versus clouds in hot Jupiters, we therefore observed Spitzer phase curves of the transiting brown dwarf KELT-1b \citep{siverd2012}. KELT-1b is a 27.23\mj\ object, with a radius of 1.116\rj. This is a mild, but significant, radius inflation compared to brown dwarf model predictions at the KELT-1 system age of 1.65\,Gyr \citep{siverd2012}. In isolation and in the field, we would expect KELT-1b to have an effective temperature of $\sim850$\,K due to its internal heat \citep{saumon2008}, but KELT-1b is on a short, 1.27 day, orbit around a 6500\,K host star. Previous observations of KELT-1b's dayside emission thus show it to be considerably hotter than the expectation from internal heat alone, at 3200\,K, and identical to a field M5 spectrum \citep{beatty2017}.

The broadband thermal dayside emission from KELT-1b has previously been observed by Spitzer at \three and \four \citep{beatty2014}, $K$ \citep{croll2015}, and $z^\prime$ \citep{siverd2012} eclipses. \cite{beatty2017} also observed an $\mathrm{R}\approx50$ $H$-band eclipse spectrum.

The relatively high mass of KELT-1b gives it a surface gravity approximately 22 times higher than a typical hot Jupiter. This high gravity could potentially change the atmospheric dynamics of KELT-1b, but since KELT-1b also receives the same level of external irradiation as a hot Jupiter, it can serve as a direct test of atmospheric dynamics in this regime.

\section{Observations and Data Reduction}

We observed two full orbit phase curves of KELT-1b at \three and \four using Spitzer/\textsc{irac}. The \three observations were taken on UT 2015 October 10, and the \four observations were taken on UT 2015 October 18. In both channels we observed for 36 hours continuously, from two hours before the ingress time of a predicted eclipse through to two hours after the egress of the succeeding eclipse. We split the observing sequence in each channel into three 12 hour long stares, so as to periodically arrest the drift of KELT-1 across the detector and recenter it onto the photometric ``sweet-spot'' at the beginning of each stare. This succeeded at \fouralt, but caused some trouble at \threealt, as we discuss below. The re-pointing process itself introduced gaps of approximately 6 minutes in between individual stares.

Our observing setup replicated that of \cite{beatty2014}. In both channels we used subarray mode with 2.0 second exposures, and PCRS peak-up mode with KELT-1 as the peak-up target to stabilize the spacecraft's pointing. We initially observed KELT-1 for 0.5 hours to allow the telescope's pointing to settle before beginning the science observations in each channel. We discarded these initial settling observations and did not use them in our analysis. In total, we collected 63,936 images at \three and \fouralt.

We began our data reduction and photometric extraction process from the basic calibrated data (BCD) images. The reduction of the KELT-1 images and the extraction of the photometry followed the process in \cite{beatty2018}, and we briefly re-describe it here.  We first determined the time of each exposure by assuming that the exposures within an individual 64-image data cube began at the \textsc{mjd\_obs} header time, and were evenly spaced between the \textsc{aintbeg} and \textsc{atimeend} header times. We converted the resulting mid-exposure times to \bjdtdb.

We next estimated the background level in each image and measured KELT-1's position. We began by masking out a box 15 pixels on a side centered on KELT-1, and taking the median of the unmasked pixels as the background level. To increase the accuracy of our background measurement, we corrected bad pixels and cosmic ray hits by performing an iterative $5\,\sigma$ clipping on the timeseries for each individual pixel and replacing outliers with the timeseries' median. The average background in our observations was $0.8\,\mathrm{e}^-\,\mathrm{pix}^{-1}$ at \three and $0.2\,\mathrm{e}^-\,\mathrm{pix}^{-1}$ at \fouralt. This was 0.07\% and 0.03\% of KELT-1's flux at \three and \fouralt, respectively. We then used the background-subtracted, bad-pixel corrected images to measure the pixel position of KELT-1 in each image using a two-dimensional Gaussian. Note that we used these corrected images only to estimate the background and to measure the position of KELT-1 -- we used uncorrected background-subtracted images for the photometric extraction.

We extracted raw photometry for KELT-1 in both channels using a circular extraction aperture centered on KELT-1's position in each image. We used an aperture radius of 2.7 pixels. For reference, the average full-width half-maximum of KELT-1's point spread function was 2.15 pixels at \three and 2.10 pixels at \fouralt. Since the fitting process for these observations was time-intensive, we did not perform a complete optimization to determine the best extraction aperture size, unlike in \cite{beatty2014} and \cite{beatty2018}. Instead, we used a fixed aperture radius of 2.7 pixels, which approximately matches the optimum aperture size for similar observations of KELT-1 determined in \cite{beatty2014}. We did perform a limited test of our aperture size by extracting and fitting photometry for aperture radii of 2.5 and 2.9 pixels. In both cases the log-likelihoods of the resulting best fits were lower, the scatter in the residuals higher, and the phase curve properties consistent with our optimum aperture at 2.7 pixels. We also tested using a variable aperture size that scaled with the noise-pixel parameter, but we found that this provided us with the same results, within the undertainties, as using a fixed aperture size.

In addition to the phase curve observations, we also re-reduced and extracted photometry from the \three and \four eclipse observations taken by \cite{beatty2014}. We used the exact same reduction and extraction process as for the phase curve observations, which added 10,880 raw images in each band to our data. As we describe below, the inclusion of the \three data from \cite{beatty2014} provided a critical bridge between the individual, 12 hour, stares in the \three phase curve data.

Finally, we trimmed outliers from the raw photometry. The first 15 minutes of the phase curve and old eclipse observations in both bands showed a clear residual ramp effect, so we excluded the first 500 points in each dataset. We removed outliers from the remaining photometry by fitting a line between the flux from the first and last point in each individual stare, and clipping those points that were more than $5\,\sigma$ away from that line. We determined the error on each point by adding the Poisson noise from KELT-1's flux in quadrature with the integrated background flux in the photometric aperture. All together, this left us with 72,162 flux measurements at \threealt, and 72,036 flux measurements at \fouralt.

\section{Lightcurve Modeling, Joint Fitting Process, and Results}

The normalized raw photometry (Figure \ref{rawplot}) showed the usual position-dependent systematics, which are caused by intra-pixel sensitivity variation in the \textsc{irac} detectors \citep[e.g.,][]{ingalls2016}. In addition, there were also discontinuous jumps in the measured flux at the beginning of each stare. These jumps were caused by slight imperfections in the PCRS peak-up process, which caused KELT-1 to begin each stare at slightly different locations on the detector (bottom two panel rows in Figure \ref{rawplot}). To correct for these effects, we used the BiLinearly-Interpolated Subpixel Sensitivity (BLISS) mapping technique \citep{stevenson2012} to simultaneously fit a subpixel sensitivity map along with an astrophysical flux model. We fit all the data in both channels simultaneously using a single set of physical parameters, but channel-dependent phase curve and BLISS parameters.

Due to KELT-1b's relatively high mass of 27.2\mj, our astrophysical flux model included several effects beyond the usual terms to describe the planetary phase variations. In particular, we also needed to account for the effect of ellipsoidal deformation and Doppler beaming on the flux from the star KELT-1 itself. In principle the high rotational velocity of KELT-1 ($v\sin(i)=56$\,km s$^{-1}$) will also cause gravity darkening on the stellar surface, but the gravity darkening coefficients for KELT-1 are both close to zero in the \textsc{irac} bandpasses. We therefore neglected rotationally-induced gravity darkening in our analysis.

The complete model we used was thus the sum of astrophysical flux changes from the star, astrophysical flux changes from the brown dwarf, and a BLISS model. Specifically, we modeled the observed flux in each band as
\begin{equation}\label{eq:3010}
F_{obs} = (F_*(t)+F_{BD}(t))\,B(x,y)\,R(r_1,t),
\end{equation}
where $B(x,y)$ is the BLISS map sensitivity for a given $x$ and $y$ pixel position of the stellar centroid, and $R(r_1,t)$ is a background linear ramp in time with slope $r_1$ to account for long term trends. The normalized flux from the star KELT-1 was
\begin{eqnarray}\label{eq:3020}
F_{*}(t) = F_{tran}(\Theta_{tran},t)&+&F_{ellip}(\Theta_{ellip},t)\\ \nonumber
&+&F_{beam}(\Theta_{beam},t),
\end{eqnarray}
where $F_{tran}(\Theta_{tran},t)$ is a transit model based on the astrophysical parameters $\Theta_{tran}$ and time $t$, $F_{ellip}(\Theta_{ellip},t)$ is a model of the stellar ellipsoidal deformation, and $F_{beam}(\Theta_{beam},t)$ represents the changes to the stellar flux from Doppler beaming.

The astrophysical model we used for the flux from the brown dwarf KELT-1b was
\begin{equation}\label{eq:3030}
F_{BD}(t) = F_{ecl}(\Theta_{ecl},t)\,\,F_{phase}(\Theta_{phase},t),
\end{equation}
where $F_{ecl}(\Theta_{ecl},t)$ is a model of the eclipse and $F_{phase}(\Theta_{phase},t)$ is a model of the phase variation, both with parameters defined similarly to above. Note that we are treating the flux from KELT-1b as the product of these two models, so that the eclipse model is normalized to be equal to zero if KELT-1b is completely behind the star and equal to one otherwise. This allows the observed eclipse depths to be set by the phase curve parameter themselves, rather than being a free parameter.

\subsection{Transit and Eclipse Models and Priors}

Both our transit and eclipse models used the \textsc{batman} Python package \citep{kreidberg2015}, which is an implementation of the \cite{mandel2002} lightcurve model. Since we normalized the eclipse lightcurve to have a depth of unity -- to allow the phase curve parameters to set the absolute eclipse depths -- both the transit and eclipse model relied upon the same set of seven physical parameters:
\begin{eqnarray}\label{eq:3110}
\Theta_{ecl} = \Theta_{tran} = && (T_C,\log P,\sqrt{e}\cos\omega,\sqrt{e}\sin\omega,\\ \nonumber
                        && \cos i, R_{BD}/R_*,\log a/R_*).
\end{eqnarray}
These were the transit center time ($T_C$), the orbital period (as $\log P$), $\sqrt{e}\cos\omega$, $\sqrt{e}\sin\omega$, the orbital inclination (as $\cos i$), the planet-to-brown dwarf radius ratio ($R_{BD}/R_*$), and the scaled semi-major axis of the orbit (as $\log a/R_*$). We calculated the secondary eclipse time using the transit center time, the orbital period, and  $\sqrt{e}\cos\omega$ and $\sqrt{e}\sin\omega$. We included a delay in the eclipse time to account for light travel time across KELT-1b's orbit, for a given value of $a/R_*$ and assuming that $R_*=1.46$\,\rsun\ \citep{siverd2012}.

\begin{deluxetable}{lcl}[b]
\tablecaption{Prior Values for KELT-1b's Properties\\ From \cite{beatty2017}}
\tablehead{\colhead{Parameter} & \colhead{Units} & \colhead{Value}}
\startdata
$T_C$\dotfill &Transit time (\bjdtdb)\dotfill & $2457306.97347\pm0.0002$\\
$P$\dotfill &Orbital period (days)\dotfill & $1.217494\pm0.000004 $\\
$\sqrt{e}\cos{\omega}$\dotfill & \dotfill & $0.005\pm0.030$\\
$\sqrt{e}\sin{\omega}$\dotfill & \dotfill & $0.004\pm0.073$\\
$\cos{i}$\dotfill &Cosine of inclination\dotfill & $0.064\pm0.019$\\
$R_{BD}/R_{*}$\dotfill &Radius ratio\dotfill & $0.07742\pm0.00056$\\
$a/R_{*}$\dotfill &Scaled semimajor axis\dotfill & $3.60\pm0.04$\\
$M_{BD}/M_{*}$\dotfill & Mass ratio\dotfill & $0.01959\pm0.0004$\\
\hline
$M_{BD}$\tablenotemark{a}\dotfill &Brown dwarf mass (\mj)\dotfill & $27.23\pm0.50$\\
$R_{BD}$\tablenotemark{a}\dotfill &Brown dwarf radius (\rj)\dotfill & $1.116\pm0.030$\\
$\log(g)_{BD}$\tablenotemark{a}\dotfill &Brown dwarf gravity (cgs)\dotfill & $4.734\pm0.025$
\enddata
\tablenotetext{a}{Not a fitting parameter, but provided for reference.}
\label{tab:priors}
\end{deluxetable}

All seven of these parameters have been measured in previous observations, and we used these independent measurements and their associated uncertainties as Gaussian priors in our fitting process. Specifically, we used the results from the $H$-band eclipse spectrum observations in \cite{beatty2017}, which we list for reference in Table \ref{tab:priors}. Note that since we used the observations from \cite{beatty2014} in our fitting, we did not include the eclipse depths measured in that paper as priors.

\subsection{Phase Curve Model and Priors}

In each channel our model for KELT-1b's phase curve variation was a single sinusoid with a variable amplitude, phase offset, and zero-point,
\begin{equation}\label{eq:3210}
F_{phase}(\Theta_{phase},t) = F_0 + C_1\,\cos\left(\frac{2\pi(t-T_C)}{P}+C_2+\pi\right),
\end{equation}
where we added $\pi$ to the quantity within the cosine so that the phase curve minimum would occur near transit for positive values of $C_1$. We also tested adding on a second harmonic at $P/2$, but as we describe in Section 3.6 we were not able to significantly detect any phase curve harmonics after the first.

The three parameters for our phase curve model were thus
\begin{equation}\label{eq:3220}
\Theta_{phase} = (T_C, \log P, F_0, C_1, C_2),
\end{equation}
where $F_0$ was the phase curve zero-point, $C_1$ the phase amplitude, and $C_2$ the phase offset. We did not impose any priors on any of these parameters, nor did we force the phase curve to stay positive around the time of transit. Though it is not included in the notation of Equation \ref{eq:3220}, we used a different set of the phase curve parameters $F_0$, $C_1$, and $C_2$ in each \textsc{irac} channel.

\subsection{Ellipsoidal Deformation and Doppler Beaming Models}

Though flux variations from ellipsoidal deformation and Doppler beaming of the host star are present in all hot Jupiter phase curve observations, the amplitude of both signals is typically small enough to be safely ignored in Spitzer observations. However, KELT-1b has a relatively high mass of 27.3\mj\ and orbits relatively close to its star at 3.6 stellar radii. As a result, we needed to account for both ellipsoidal deformation and Doppler beaming -- in addition to the typical phase curve terms.

\begin{figure}[t]
\vskip 0.00in
\includegraphics[width=1.0\linewidth,clip]{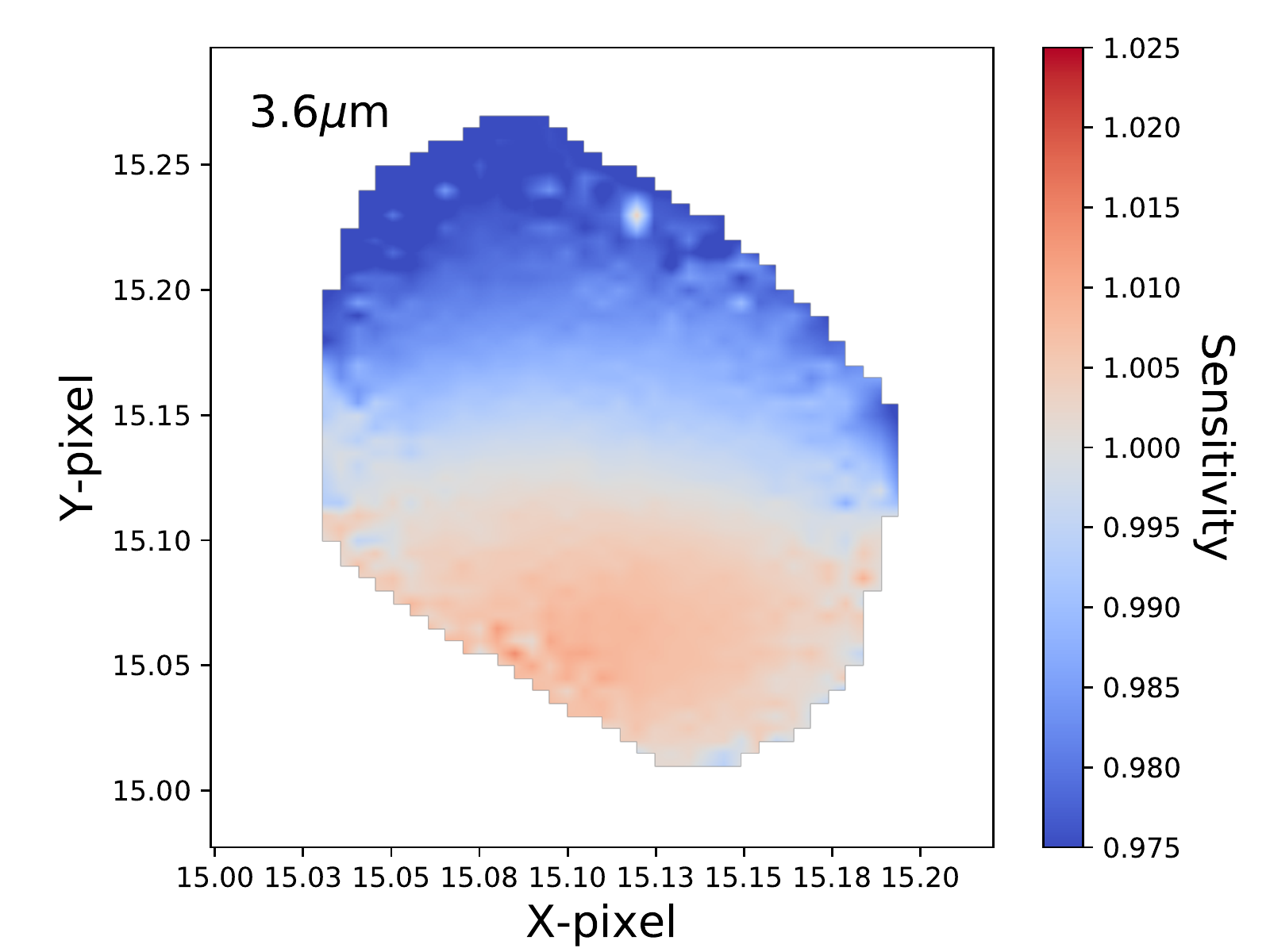}
\vskip -0.0in
\caption{BLISS sensitivity map for our \three observations, as described in Section 3.4.}
\label{blissgrid36}
\end{figure}

Since our transit and eclipse models account for orbital eccentricity, orientation, and inclination, we calculated the ellipsoidal deformation signal using an expanded analytic form that allows for an eccentric and inclined orbit:
\begin{equation}\label{eq:3310}
F_{ellip}(\Theta_{ellip},t) = A_{ellip}\,(1-\cos\left[\frac{4\pi(t-T_C)}{P}\right]),
\end{equation}
where $T_C$ and $P$ are defined as in Section 3.1. The amplitude, $A_{ellip}$, of the ellipsoidal deformation signal was
\begin{equation}\label{eq:3320}
A_{ellip} = \beta\, \frac{M_{BD}}{M_*}\left(\frac{R_*}{a}\right)^3\left(\frac{1+e\cos\nu}{1-e^2}\right)^3\,\sin(i)^3.
\end{equation}
In this equation $\nu$ was the true anomaly of KELT-1b along its orbit at time $t$, and the final two terms in the equation accounted for the changing brown dwarf-star separation in a possibly eccentric orbit, and the orbital inclination, respectively. $\beta$ was a gravity darkening term, which we estimated following \cite{mazeh2010} as
\begin{equation}\label{eq:3330}
\beta = 0.12 \frac{(15+u)(1+g)}{3-u},
\end{equation}
where $g$ is the linear stellar gravity darkening coefficient and $u$ is the linear stellar limb-darkening coefficient. According to \cite{claret2011} these coefficients are approximately the same for the star KELT-1 in the Spitzer bandpasses, so in both channels we fixed $g=0.08$ and $u=0.15$, for $\beta=0.68$.

\begin{figure}[t]
\vskip 0.00in
\includegraphics[width=1.0\linewidth,clip]{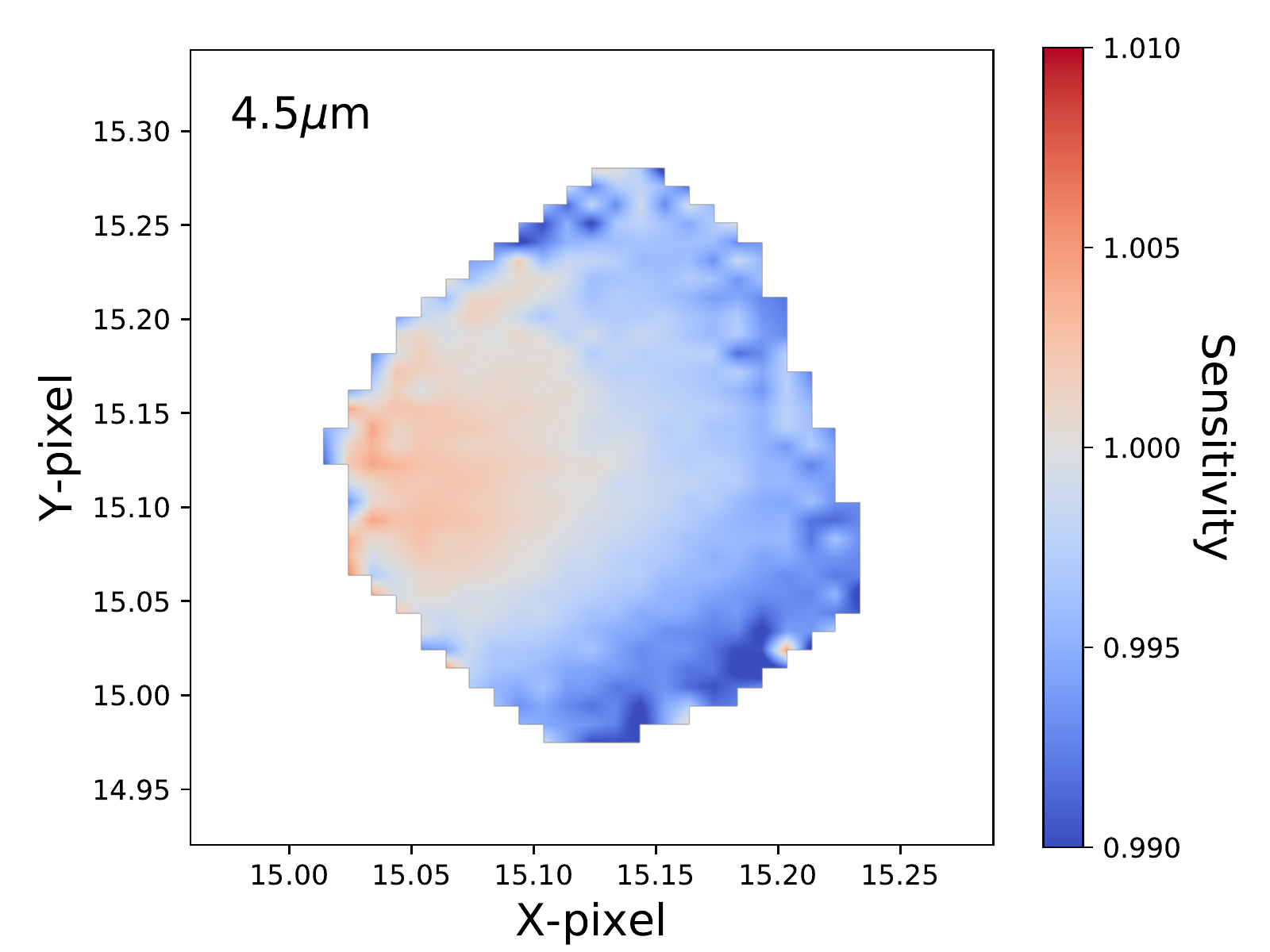}
\vskip -0.0in
\caption{BLISS sensitivity map for our \four observations, as described in Section 3.4.}
\label{blissgrid45}
\end{figure}

All together the parameters determining the ellipsoidal deformation model were
\begin{eqnarray}\label{eq:3335}
\Theta_{ellip} = &&(T_C, \log P, M_{BD}/M_*, \log a/R_*, \\ \nonumber
                          && \sqrt{e}\cos\omega, \sqrt{e}\sin\omega, \cos i).
\end{eqnarray}
Note that the argument of periastron, $\omega$, came into Equation \ref{eq:3320} during the calculation of the true anomaly, $\nu$.

To calculate the Doppler beaming signal we followed \cite{loeb2003}, such that
\begin{equation}\label{eq:3340}
F_{beam}(\Theta_{beam},t) =(3-\alpha)\frac{K_{RV}(t)}{c},
\end{equation}
where $c$ was the speed of light and $K_{RV}$ was the stellar radial velocity at time $t$. We calculated $\alpha$ using Equation 3 from \cite{loeb2003} using a 6500\,K blackbody for the spectrum of the star KELT-1 in the Spitzer bandpasses. We found that $\alpha$ was effectively the same in both channels, and so used a single value for both. To allow for eccentric and inclined orbits we calculated the stellar radial velocity as
\begin{equation}\label{eq:3350}
K_{RV}(t) = \frac{2\pi\,a_*\sin i}{P\,\sqrt{1-e^2}}\,\cos(\omega+\nu_*)+e\,\cos\omega,
\end{equation}
where $a_*$ was the semi-major axis of the star KELT-1's orbit, and $\nu_*$ was the true anomaly of KELT-1. We calculated $a_*$ as 
\begin{equation}\label{eq:3360}
a_* = \left(\frac{M_{BD}}{M_*}\right)\left(\frac{a}{R_*}\right)R_*,
\end{equation}
and assumed that $R_*=1.46$\,\rsun. The parameters determining the Doppler beaming signal were then
\begin{eqnarray}\label{eq:3270}
\Theta_{beam} = &&(T_C, \log P, M_{BD}/M_*, \log a/R_*, \\ \nonumber
                            && \sqrt{e}\cos\omega, \sqrt{e}\sin\omega, \cos i).
\end{eqnarray}

\subsection{BLISS Model and Ramp}

As can be seen in Figure \ref{rawplot}, the photometry in both channels showed clear correlations to the $x$ and $y$ pixel position of KELT-1b on the detectors. These correlations are typical in Spitzer/\textsc{irac} timeseries photometry, and are primarily a result of intra-pixel sensitivity variations on the \textsc{irac} detectors. Many different techniques have been used over the years to account for these intra-pixel effects. For these observations we chose to use the BLISS mapping technique described by \cite{stevenson2012}.

BLISS mapping attempts to fit the detector's intra-pixel sensitivity variations simultaneously with the astrophysical signal. It does so by taking the residuals between the observed flux ($F_{obs}$) and a proposed astrophysical flux model ($F_*+F_{BD}$) and assuming that those residuals are primarily caused by intra-pixel affects. Using the measured $x$ and $y$ pixel positions of the stellar centroid, BLISS mapping then models the detector sensitivity by constructing a bilinear interpolation of the flux residuals as a function of $x$ and  $y$ pixel position. Thus for a given astrophysical model
\begin{equation}\label{eq:3410}
B(x,y) = \frac{F_{obs}}{(F_*+F_{BD})\,R(r_1,t)}.
\end{equation}
Following \cite{stevenson2012}, we also included a linear ramp term, $R(r_1,t)$ in our BLISS model. This had the form
\begin{equation}\label{eq:3420}
R(r_1,t) = r_1\,(t-\tilde{t}) + 1,
\end{equation}
where $\tilde{t}$ was the median observation time in a single stare in a single channel, and the slope $r_1$ was a free parameter. While the BLISS map $B(x,y)$ was shared between all the observations in an individual channel, each stare within a channel had its own individual slope parameter. To construct each channel's BLISS map we first needed to specify a rectangular grid of interpolation ``knots,'' which determined the resolution of the resulting sensitivity map. As noted in \cite{stevenson2012}, too fine an interpolation grid results in the BLISS map fitting the observing noise itself, while too coarse a grid will fail to totally account for intra-pixel sensitivity variations. Following the recommendation in \cite{stevenson2012} we set the grid spacing on both axes of both channels to be 0.015 pixels. This spacing also happened to be approximately three times the standard deviation of the frame-to-frame differences in the x- and y-positions of the stellar centroid, which means that each observation should be correctly associated with its nearest grid point.

Recall that in each \textsc{irac} channel we had three individual 12 hour stares within the phase curve observations, and an additional stare covering a secondary eclipse from \cite{beatty2014}. As just said, we constructed a single unified BLISS map by combining all four stares in each channel (Figures \ref{blissgrid36} and \ref{blissgrid45}), but we allowed the slope of the linear ramp to vary between stares. Additionally, for the \cite{beatty2014} observations we included a floating offset in the ramp term, to account for flux normalization differences between those data and the phase curve data.

In describing the phase curve observations, we mentioned that splitting these observations into three individual stares caused some initial problems in our fitting. The practice of splitting up long duration, continuous, photometric observations into 12 hour long stares is recommended by the Spitzer Science Center, since it causes the spacecraft to periodically reacquire and recenter the target star. The idea is to prevent the target star drifting off the ``sweet-spot'' on the \textsc{irac} detectors that is well-characterized for photometric observations. This worked at \fouralt, but caused significant problems at \threealt.

The issue in the \three observations was that the second and third stares in the phase curve observations recentered KELT-1 approximately 0.04 pixels to the left of the first stare (Figure \ref{rawplot}). While small, this offset meant that there was effectively no overlap between the stellar centroid positions during the first \three stare and the subsequent \three stares. As a direct result of this, we found it impossible to construct a unified intra-pixel sensitivity map for the \three observations using BLISS mapping, Gaussian Process regression, or pixel-level decorrelation (PLD): the photometry in the first \three stare always had a floating offset relative to the second and third stares. This made it impossible to accurately or precisely determine the phase curve parameters. Luckily, the old \three observations from \cite{beatty2014} were positioned on the detector such that they managed to bridge the gap between the phase curve stares. Including these data in our fitting stabilized the fitted sensitivity maps, and allowed us to measure KELT-1b's phase curve.  

\begin{figure*}[t]
\vskip 0.00in
\includegraphics[width=1.05\linewidth,clip]{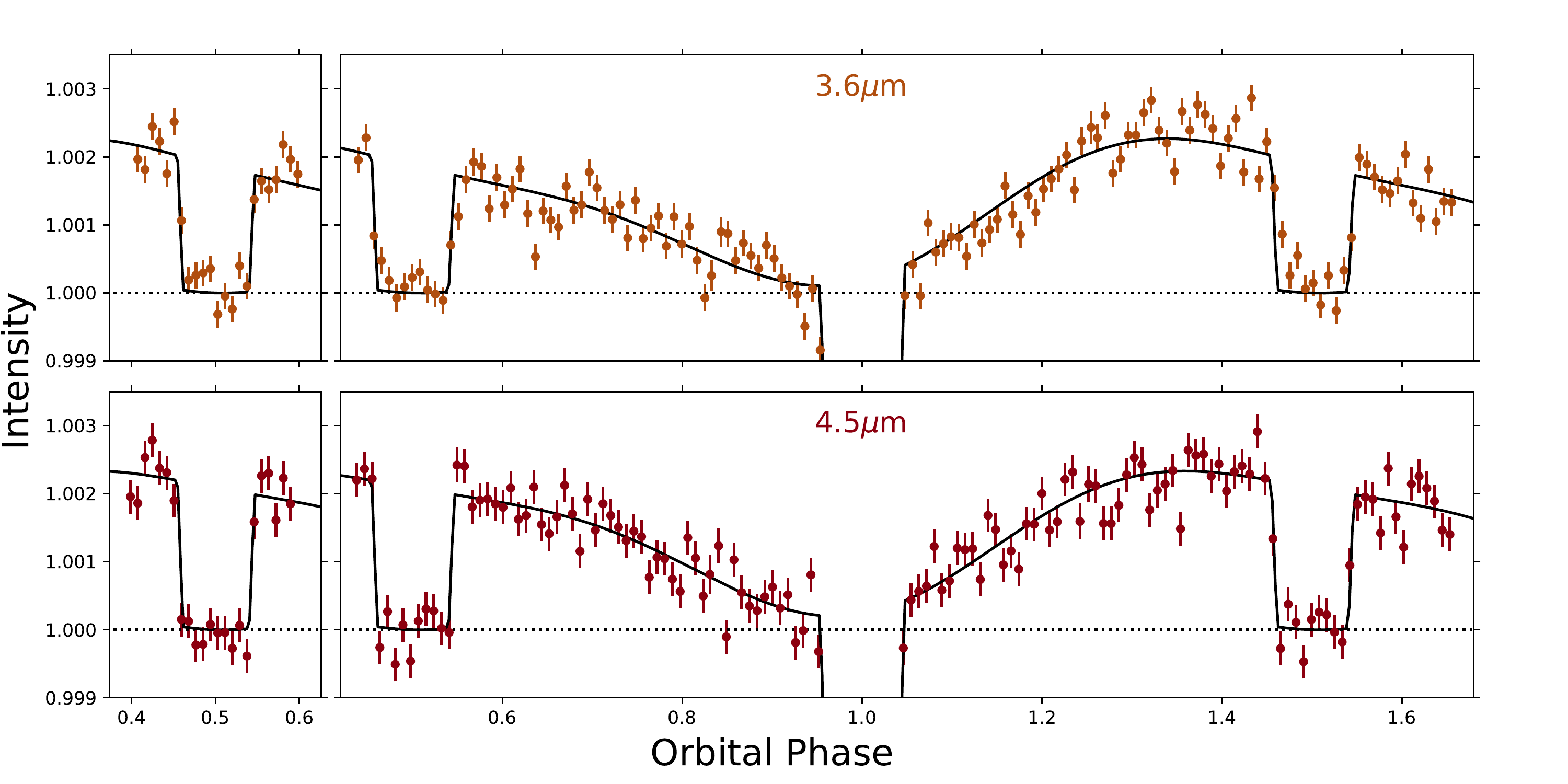}
\vskip -0.0in
\caption{The detrended phase curve and eclipse photometry for KELT-1b, as determined via the joint fitting process described in Section 3.5. The narrow panels on the left show the refit eclipse data from 2012, and the broad panels on the right show the new phase curve data. Due to KELT-1b's relatively high mass the apparent phase variation in this detrended photometry contains significant amounts of signal from stellar ellipsoidal deformation (250\,ppm) and Doppler beaming (56\,ppm), in addition to signal from the planetary phase variation ($\approx970$\,ppm). See Figure \ref{phaseelements} for a decomposition of these signals.}
\label{detrend}
\end{figure*}

There are two takeaways from this. First, \textsc{irac} observations of the same target using the same observing setup are stable relative to each other for at least three years (i.e., the time between the \cite{beatty2014} observations and these phase curve observations). Second, observers should realize that splitting long duration Spitzer observations up into individual stares carries the risk that the repointing process will randomly recenter the target star onto a completely different section of the detector. Without some method of bridging the observations at different positions, this will cause problems in fitting the results. Indeed, this exact problem occurred in \cite{stevenson2017}'s \three phase curve observations of WASP-43b, which the authors solved by taking completely new \three observations. Alternately, \cite{mendonca2018} were able to determine a set of ``stable'' phase curve parameters from these WASP43b data by using a combination of BLISS mapping and decorrelation against the FWHM of the stellar PSF \citep{lanotte2014}. From the analysis in \cite{mendonca2018}, this analysis method appears robust to the problems caused by BLISS mapping islands.

Finally, we note that even with the inclusion of the \cite{beatty2014} observations neither a Gaussian Process (GP) model nor a PLD model converged to a stable offset between the first \three stare and the two following observing sequences. Given the success of our BLISS mapping fits we did not determine the exact cause of these two fitting failures, but on the surface this indicates that BLISS mapping is more robust to observing issues in phase curve time-series than a GP- or PLD-based detrend. Fortunately, in our case these failures presented themselves as extremely large uncertainties from the two analyses, rather than giving us apparently precise -- but inaccurate -- results.

\subsection{Joint Fitting Process}

We simultaneously fit the phase curve and eclipse observations in both bands. To do so, we used a single set of stellar, brown dwarf, and orbital parameters for all four datasets (one phase curve and one eclipse, in two bands), and two separate sets of phase curve parameters at \three and \fouralt. Recall that we imposed Gaussian priors as listed for the parameters in Table \ref{tab:priors}, and that we used the same gravity-darkening and Doppler beaming coefficients for both the \three and \four observations. We imposed no priors on the phase curve parameters, nor did we require that the minimum flux in the phase curve be always greater than zero. 

\begin{figure*}[t]
\vskip 0.00in
\includegraphics[width=1.05\linewidth,clip]{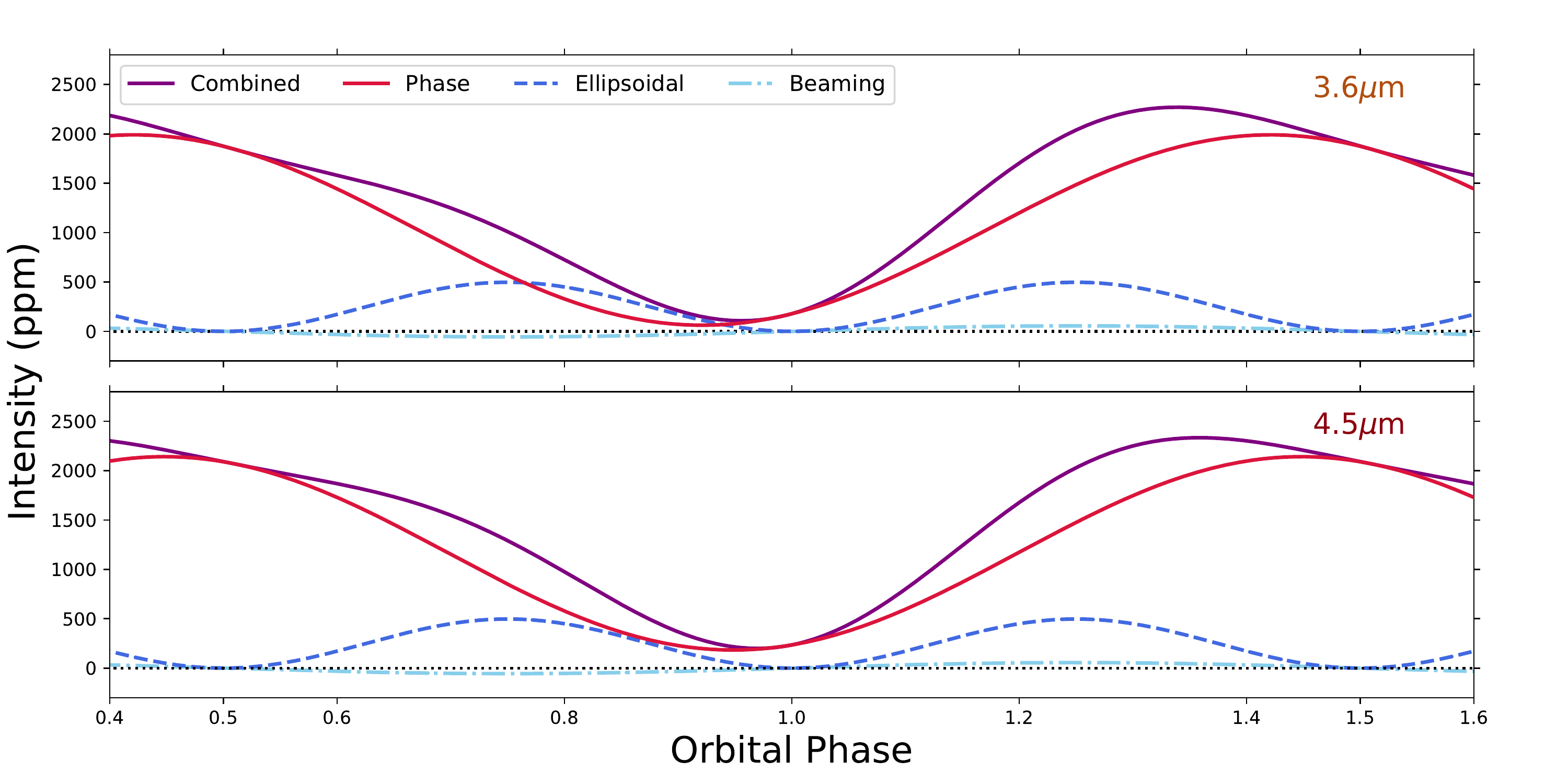}
\vskip -0.0in
\caption{Our phase curve observations of KELT-1b show signatures of stellar ellipsoidal deformation and Doppler beaming, as described in Section 3.3. This makes the combined, observed, phase curve in each Spitzer band (purple line) appear noticeably non-sinusoidal, even though the planetary phase variation is best fit using a single sinusoidal harmonic (red line, and end of Section 3.6).}
\label{phaseelements}
\end{figure*}

We began the fitting process by conducting a Nelder-Mead likelihood maximization to identify an initial best fit. We then used an \textsc{mcmc} analysis to explore parameter space around this initial best fit to determine the global maximum likelihood and estimate parameter uncertainties. To do the \textsc{mcmc} analysis, we used the \emph{emcee} Python package \citep{dfm2013} to run 60 walkers with a 3,000 step burn-in followed by a 30,000 step production sequence. We initialized the walkers in a Gaussian ball about the Nelder-Mead fit location. To judge the convergence of the \textsc{mcmc} chains we used the Gelman-Rubin (GR) test statistic, which we required to be below 1.05 for each parameter. We do note that because \emph{emcee} uses an Affine Invariant sampler, the GR statistic is not a perfect convergence test since it presumes that each \textsc{mcmc} chain is perfectly independent. As an alternate convergence metric, we also calculated the autocorrelation lengths for each parameter, which varied from approximately 400 to 600. Since the autocorrelation lengths were all less than or equal to 1/50 of the total number of \textsc{mcmc} steps, we considered this good evidence for convergence. 

\subsection{Results}

The results from the joint fitting of the data are listed in Table 2 and shown in Figures \ref{detrend} and \ref{phaseelements}. We clearly detect phase variation from KELT-1b in both bands, as $C_{1,3.6}=959\pm41$\,ppm and $C_{1,4.5}=969\pm51$\,ppm. We also measure a significant Eastward offset of the phase maximum, of $C_{2,3.6}=28.6\pm3.8$ degrees and $C_{2,4.5}=18.5\pm5.1$ degrees. The eclipse depths we measure are $\delta_{3.6}=1877\pm58$\,ppm and $\delta_{4.5}=2083\pm70$\,ppm. These depths are consistent with the eclipse depths measured by \cite{beatty2014}, though note that we are including the \cite{beatty2014} data in our analysis.

Note that in Table 2 both the ellipsoidal deformation and Doppler beaming amplitudes are listed as ``derived'' parameters, since we did not fit for them independently. Instead, we calculated both amplitudes using the relations described in Section 3.3. This means that the uncertainties for the ellipsoidal deformation and Doppler beaming amplitudes listed in Table 2 are dominated by our priors on the fit parameters listed in Equations \ref{eq:3335} and \ref{eq:3270}.

Though our best fit results are for a phase curve composed of a single sinusoidal harmonic, we experimented with adding higher order harmonics. In particular, a phase curve model with an additional second order harmonic of $C_3\,\cos([2\pi(t-T_C)/2P]+C_4+\pi)$ provided a noticeably higher maximum likelihood than the single harmonic model, of $\Delta \ln p = 13.4$. However, the Bayesian Information Criteria (BIC) of the two phase curve fits strongly preferred the model with only the single harmonic, at $\Delta \mathrm{BIC}=-20.7$. As a result, we adopted the single harmonic phase curve model as the best fit to the data.

\begin{deluxetable*}{lccc}
\tablecaption{Median Values and 68\% Confidence Intervals for the Joint \three and \four Fit}
\tablehead{\colhead{~~~Parameter} & \colhead{Units} & \multicolumn{2}{c}{Value}}
\startdata               
\sidehead{Joint Parameters:}
~~~$T_C$\dotfill &Transit time (\bjdtdb)\dotfill & \multicolumn{2}{c}{$2457306.97602\pm0.0003$}\\
~~~$\log(P)$\dotfill &Log orbital period (days)\dotfill & \multicolumn{2}{c}{$0.0854664\pm2\times10^{-7}$}\\
~~~$\sqrt{e}\cos{\omega}$\dotfill & \dotfill &\multicolumn{2}{c}{$0.02\pm0.03$}\\
~~~$\sqrt{e}\sin{\omega}$\dotfill & \dotfill & \multicolumn{2}{c}{$-0.002\pm0.025$}\\       
~~~$\cos{i}$\dotfill & Cosine of inclination\dotfill & \multicolumn{2}{c}{$0.054\pm0.015$}\\
~~~$R_{BD}/R_{*}$\dotfill &Radius ratio\dotfill & \multicolumn{2}{c}{$0.0771\pm0.0003$}\\
~~~$\log(a/R_{*})$\dotfill &Log semi-major axis in stellar radii\dotfill & \multicolumn{2}{c}{$0.568\pm0.004$}\\
~~~$M_{BD}/RM_{*}$\dotfill &Mass ratio\dotfill & \multicolumn{2}{c}{$0.01958\pm0.0004$}\\[1ex]
Phase Curve Parameters: & & \multicolumn{1}{c}{\underline{\textbf{\three}}} &  \multicolumn{1}{c}{\underline{\textbf{\four}}} \\
~~~$F_{0}$\dotfill & Phase baseline (ppm)\dotfill & $1037\pm39$ &  $1161\pm48$\\
~~~$C_{1}$\dotfill & Phase amplitude (ppm)\dotfill & $959\pm41$ &  $976\pm51$\\
~~~$C_{2}$\dotfill & Phase offset (deg.)\dotfill & $28.6\pm3.8$ &  $18.5\pm5.1$\\
\sidehead{BLISS Model Parameters:}
~~~$r_{1}$\dotfill & Linear ramp, stare 1\dotfill & $0.0070\pm0.0004$ &  $0.0014\pm0.0007$\\
~~~$r_{2}$\dotfill & Linear ramp, stare 2\dotfill & $0.0066\pm0.0004$ &  $-0.0027\pm0.0008$\\
~~~$r_{3}$\dotfill & Linear ramp, stare 3\dotfill & $0.0057\pm0.0007$ &  $0.0012\pm0.0011$\\
~~~$r_{2012}$\dotfill & Linear ramp, 2012 data\dotfill & $0.0024\pm0.0008$ &  $0.0030\pm0.0010$\\
~~~$c_{2012}$\dotfill & Normalization offset, 2012 data\dotfill & $0.00524\pm0.00007$ &  $0.0019\pm0.0001$\\[1ex]
\hline\hline \\ [-2ex]
Derived Phase Curve Parameters: & & \multicolumn{1}{c}{\underline{\textbf{\three}}} &  \multicolumn{1}{c}{\underline{\textbf{\four}}} \\
~~~$\delta$\dotfill & Secondary eclipse depth (ppm)\dotfill & $1877\pm58$ &  $2083\pm70$\\
~~~$F_{night}$\dotfill & Nightside flux (ppm)\dotfill & $197\pm52$ &  $238\pm64$\\
~~~$F_{max}$\dotfill & Phase maximum (ppm)\dotfill & $1994\pm56$ &  $2133\pm69$\\
~~~$F_{min}$\dotfill & Phase minimum (ppm)\dotfill & $81\pm35$ &  $188\pm68$\\
\sidehead{Derived Joint Parameters:}
~~~$A_{ellip}$\dotfill & Ellipsoidal def. amplitude (ppm)\dotfill & \multicolumn{2}{c}{$250\pm86$}\\
~~~$A_{beam}$\dotfill & Dopp. beaming amplitude (ppm)\dotfill & \multicolumn{2}{c}{$56\pm1$}\\
~~~$T_S$\dotfill &Secondary eclipse time (\bjdtdb)\dotfill & \multicolumn{2}{c}{$2457307.58552\pm 0.00033$}\\
~~~$P$\dotfill &Orbital period (days)\dotfill & \multicolumn{2}{c}{$1.2174928\pm6\times10^{-7}$}\\  
~~~$i$\dotfill & Inclination (deg.)\dotfill & \multicolumn{2}{c}{$86.8\pm0.8$}\\
~~~$a/R_{*}$\dotfill &Semi-major axis in stellar radii\dotfill & \multicolumn{2}{c}{$3.693\pm0.038$}\\
~~~$b$\dotfill &Impact parameter\dotfill & \multicolumn{2}{c}{$0.21\pm0.05$}\\
~~~$e$\dotfill & Eccentricity\dotfill &\multicolumn{2}{c}{$0.0013\pm0.0005$}\\
~~~$\omega$\dotfill & Argument of Periastron\dotfill & \multicolumn{2}{c}{$358\pm51$}
\enddata
\label{tab:results}
\end{deluxetable*}

\section{Observable Signatures of Clouds in Spitzer Phase Curve Measurements}

Our measurements of the phase offset and the phase amplitude in both bands are generally what one would expect based on Spitzer phase curve observations of hot Jupiters. KELT-1b's phase offsets are higher than those of similarly irradiated planets by roughly $3\sigma$ to $5\sigma$, but this may be consistent with the high scatter seen in other Spitzer phase offsets. The phase amplitudes we measure for KELT-1b are in the center of the observed distribution, and follow the trend from hot Jupiters that day-night temperature contrasts increase with increasing irradiation \citep{showman2002,perezbecker2013}.

One key assumption in the above is that the longitudinal temperature distribution we see in KELT-1b's atmosphere is set by thermal and dynamical processes. That is, modeling the atmosphere is fundamentally an energy transport problem governed by the ratio of either the radiative and advective timescales \citep{showman2008}, or the radiative and drag timescales \citep{komacek2016}.

However, it seems more likely that KELT-1b's observed thermal emission is being heavily modulated by clouds. Furthermore, when we consider the ensemble phase curve properties of KELT-1b and the hot Jupiters, it appears that essentially all of the objects for which we have \three and \four Spitzer phase curves also show evidence for clouds affecting their thermal emission properties.

We see this in four different ways:
\begin{enumerate}
\item The most recent set of published phase offset measurements -- including KELT-1b -- are consistent with the observed planets having a constant phase offset of 14 deg. for all planetary equilibrium temperatures, though with a high scatter. This conflicts with thermal-only global circulation model (GCM) predictions that cooler planets should show large ($\sim70$\,deg.) offsets \citep{zhang2018}.
\item The relatively low nightside flux that we measure for KELT-1b requires that the underlying atmospheric intensity map (as opposed to the disk-integrated flux we actually observe) display a sharp transition between the day- and nightsides, and be at a roughly constant intensity level across the nightside.
\item When we examine the day- and nightside brightness temperatures for all the planets at \three and \fouralt, we see two remarkable trends. First, the dayside brightness temperatures show a strong linear trend as a function of planetary equilibrium temperature. Second, the nightside brightness temperatures in both bands are consistent with all the planets having constant, $\sim1000$\,K, nightsides.
\item Using Gaia DR2 parallaxes for KELT-1b and the other planets we can trace the phase evolution of their atmospheres on a color-magnitude diagram. These trajectory plots suggest that the planets have nightside colors that are only explained by the presence of clouds.
\end{enumerate}

For reference, the planets that we compared KELT-1b to in the rest of this section were all the hot Jupiters which have published \three or \four Spitzer phase curve observations. These were: HAT-P-7b \citep{wong2016}, HD 189733b \citep{knutson2012}, HD 149026b \citep{zhang2018}, WASP-12b \citep{cowan2012}, WASP-14b \citep{wong2015}, WASP-18b \citep{maxted2013}, WASP-19b \citep{wong2016}, WASP-33b \cite{zhang2018}, WASP-43b \citep{mendonca2018}, and WASP-103b \citep{kreidberg2018}. We also included the \four results for HD 209458b \citep{zellem2014}. Note that we used the recent \cite{mendonca2018} reanalysis of the WASP-43b phase curves, rather than the original \cite{stevenson2017} results. Note too that we did not include HAT-P-2b \citep{lewis2013}, since it is on a very eccentric ($e=0.52$) orbit and therefore will have different atmospheric dynamics compared to the other hot Jupiters on circular orbits.

\subsection{Phase Offsets and Their Lack of Variation With Temperature}

Thermal-only atmospheric models generally predict a strong correlation between the zero-albedo planetary equilibrium temperature, and the measured Eastward phase offset in the planetary atmosphere. As described in \cite{zhang2018}, cloudless GCMs from \cite{kataria2016} predict a trend with equilibrium temperature of $m_{\mathrm{\textsc{gcm}}}\approx-0.04\,\mathrm{deg.}\,\mathrm{K}^{-1}$, which would imply a 70 deg. difference between the offsets of the hottest and the coolest planets. Similarly, simple thermal-only energy transport models \citep{cowan2011,schwartz2017} also indicate that planetary atmospheres that zonally advect heat from the day- to nightside should show higher phase curve offsets at lower equilibrium temperatures.

Observationally, no strong correlation between phase offsets and day-night temperature contrast has been seen, which has been considered indicative of non-thermal influences on the hotspots locations \citep[e.g.][and references therein]{crossfield2015}. Recently, \cite{zhang2018} noted an apparent two-part trend for phase offsets as a function of the zero-albedo planetary equilibrium temperature\footnote{\cite{zhang2018} actually used the planetary ``irradiation'' temperature, which is $T_{ir} = \sqrt{2}\,T_{eq}$. For this discussion we have converted their results to $T_{eq}$.}. The authors suggested that this two-part trend could be caused by the increasing amounts of high-altitude dayside clouds up to 2400K, followed by dispersal of the clouds in the planets hotter than the breakpoint.  

\begin{figure}[t]
\vskip 0.00in
\includegraphics[width=1.05\linewidth,clip]{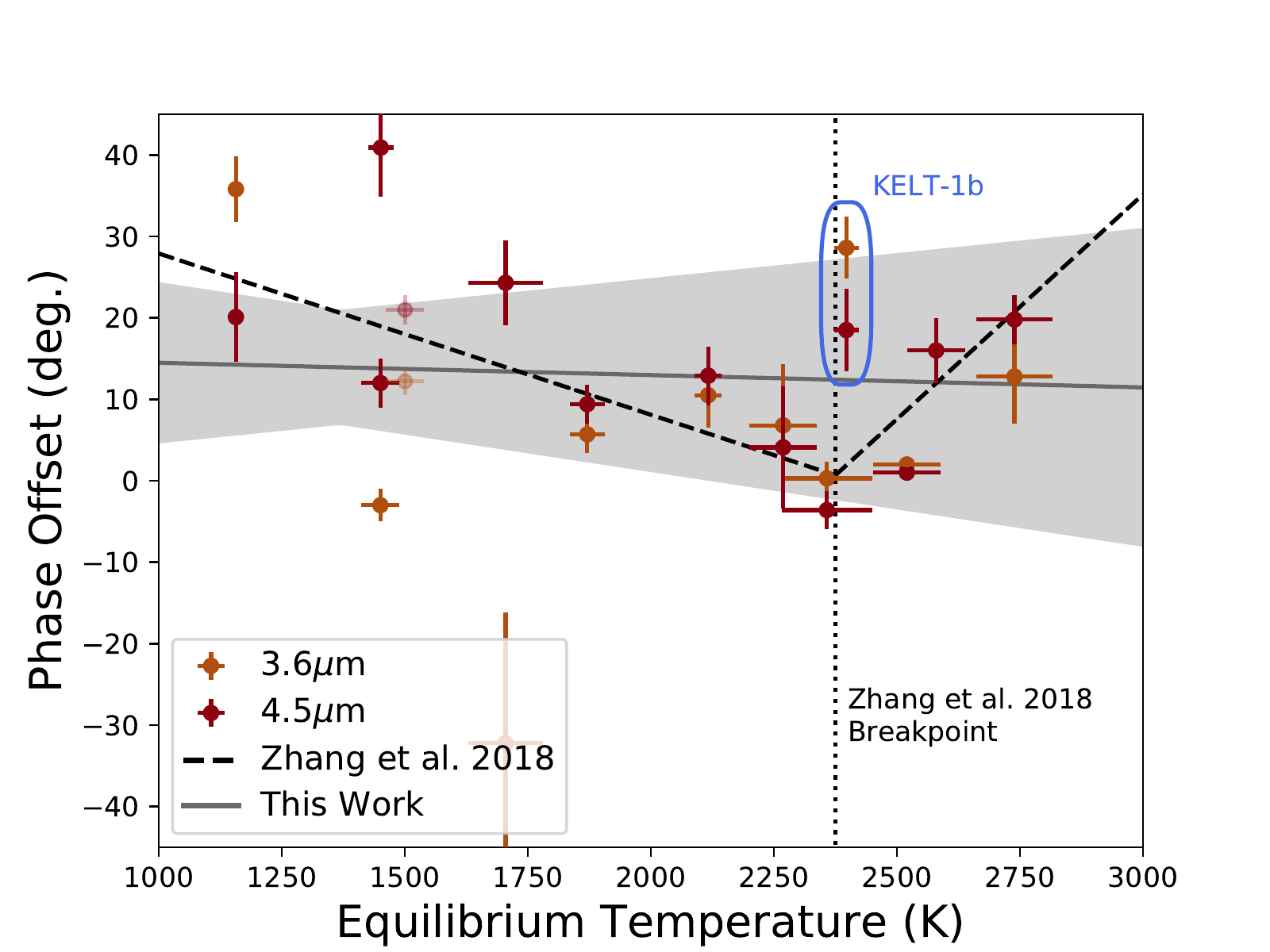}
\vskip -0.0in
\caption{\three and \four Phase curve offsets as a function of equilibrium temperature. The light blue rectangle shows our measured offsets for KELT-1b. For our analysis we used the \cite{mendonca2018} results for WASP-43b, rather than the \cite{stevenson2017} results, but as a comparison we also show the \cite{stevenson2017} points as the two greyed out points offset just to the right of the \cite{mendonca2018} offsets. Recently \cite{zhang2018} suggested that Spitzer phase curve offsets follow a two stage trend: linearly decreasing up to $T_{eq}=2400$\,K, and linearly increasing thereafter (black dashed line). However, while trying to replicate \cite{zhang2018}'s results we discovered an undocumented numerical effect in Python's \emph{scipy.odr} package that causes unrealistically small uncertainties in linear regressions (see the discussion in Section 4.1). When we correct for this effect we find that that the offsets of the cooler planets do not show a strong trend with temperature. Instead, the simplest explanation of these offset measurements is that they are approximately constant with temperature, but with a high scatter: we find a combined slope of  $m_{\mathrm{all}}=-0.002\pm0.008\,\mathrm{deg.}\,\mathrm{K}^{-1}$ (grey line with the shaded region as the $1\,\sigma$ uncertainty).}
\label{offsets}
\end{figure}

\cite{zhang2018} specifically found that the combined \three and \four phase offsets for planets cooler than 2400K decreased as the equilibrium temperature increases at a rate of $-0.020\pm0.003\,\mathrm{deg.}\,\mathrm{K}^{-1}$. Planets hotter than the 2400K breakpoint then began to show increasing phase offsets at the rate of $0.055^{+0.024}_{-0.016}\,\mathrm{deg.}\,\mathrm{K}^{-1}$ (dashed black line, Figure \ref{offsets}). \cite{zhang2018} did caution that the trend for the hotter planets was based on only three observations, and might not be real.

One extremely important point that we discovered while attempting to replicate \cite{zhang2018}'s result is a numerical failure in Python's \emph{scipy.odr} package. This package is a wrapper for the \textsc{odrpack} routines from \textsc{fortran 77}, and is intended to fit functions to data with uncertainties in both the x- and y-directions. Using \emph{scipy.odr}, we were able to perfectly replicate \cite{zhang2018}'s measured slopes and slope uncertainties. Unfortunately, we discovered that \emph{scipy.odr} only weights data points by their \emph{relative} uncertainties, not the absolute uncertainties. Practically, this means that if one were to multiply all of the temperature and offset uncertainties shown in Figure \ref{offsets} by, for example, a factor of 10, then \emph{scipy.odr} would still return the same slope and slope uncertainty as for the original data. As a result, regressions done using \emph{scipy.odr} give unrealistically small uncertainties. Note that this behavior does not appear to be mentioned in the package documentation.

We instead switched to using the \emph{BCES} Python package, which is based on the \cite{bces} method for conducting a linear regression on data with measurement uncertainties in both the x- and y-directions. This new regression does not quite replicate \cite{zhang2018}'s result. Instead, for the same data set we find that the phase offsets in planets cooler than 2400K decrease at a rate of $-0.018\pm0.007\,\mathrm{deg.}\,\mathrm{K}^{-1}$. This is only a marginally significant trend in phase offsets for these cooler planets.

In addition to trying to replicate \cite{zhang2018}'s findings, we also added our measurement of KELT-1b's phase offsets and the phase offsets from \cite{kreidberg2018}'s recent observations of WASP-103b ($T_{eq}=2520$\,K). Both data sets are for planets with $T_{eq}>2400$\,K, and while the WASP-103b offsets are roughly consistent with \cite{zhang2018}'s suggested trend as a function of temperature, KELT-1b's phase offsets are considerably higher than predicted by the \cite{zhang2018} trend. Indeed, with the addition of these two planets we find that phase offsets above  $T_{eq}>2400$\,K roughly decrease with increasing temperature, though the slope of this trend is not significant, at $m_{\mathrm{hot}}=-0.005\pm0.02\,\mathrm{deg.}\,\mathrm{K}^{-1}$.

For the cooler planets below $T_{eq}<2400$\,K, we made two additional changes to the \cite{zhang2018} analysis. First, we used the recent re-analysis of WASP-43b's ($T_{eq}=1450$\,K) phase curve conducted by \cite{mendonca2018}, rather than the results from \cite{stevenson2017} used in \cite{zhang2018}. The \cite{mendonca2018} analysis found that WASP-43b's phase offsets in both bands were approximately 10 degrees lower than \cite{stevenson2017}'s analysis of the same data. Second, we added the \four measurement of HD 209458b from \cite{zellem2014}. These did not change the new slope of the offsets for the cooler planets we measured above: we again find a marginally significant slope of $m_{\mathrm{cool}}=-0.018\pm0.007\,\mathrm{deg.}\,\mathrm{K}^{-1}$ (Figure \ref{offsets}).

Rather than a two-part trend in phase offset vs. temperature, the simplest explanation of the observations appears to be that phase offsets are constant as a function of equilibrium temperature -- albeit with a high scatter. A straight line fit to all of the points in Figure \ref{offsets} gives a slope consistent with zero, of $m_{\mathrm{all}}=-0.002\pm0.008\,\mathrm{deg.}\,\mathrm{K}^{-1}$. If we then fit the observations assuming a constant phase offset of $14$\,deg., this constant fit has a lower BIC than either a single sloped line ($\Delta\mathrm{BIC}=2.3$), or a two-part fit ($\Delta\mathrm{BIC}=3.1$). We do note, however, that these differences are only mildly significant.

The constant phase offsets as a function of temperature, or at least the lack of a clear trend conflicts with the previously mentioned predictions from GCMs and basic thermal transport models that offsets should be large ($\sim70$ deg.) at lower temperatures. As mentioned in those modeling papers, the likely culprit is clouds \citep{showman2013,komacek2016}. Specifically, the observed offsets for the cooler planets are being suppressed to lower values (i.e., closer to planetary noon) than predicted by purely thermal models due to cloud formation during planetary dusk and late afternoon.

\begin{figure*}[t]
\vskip 0.00in
\includegraphics[width=1.05\linewidth,clip]{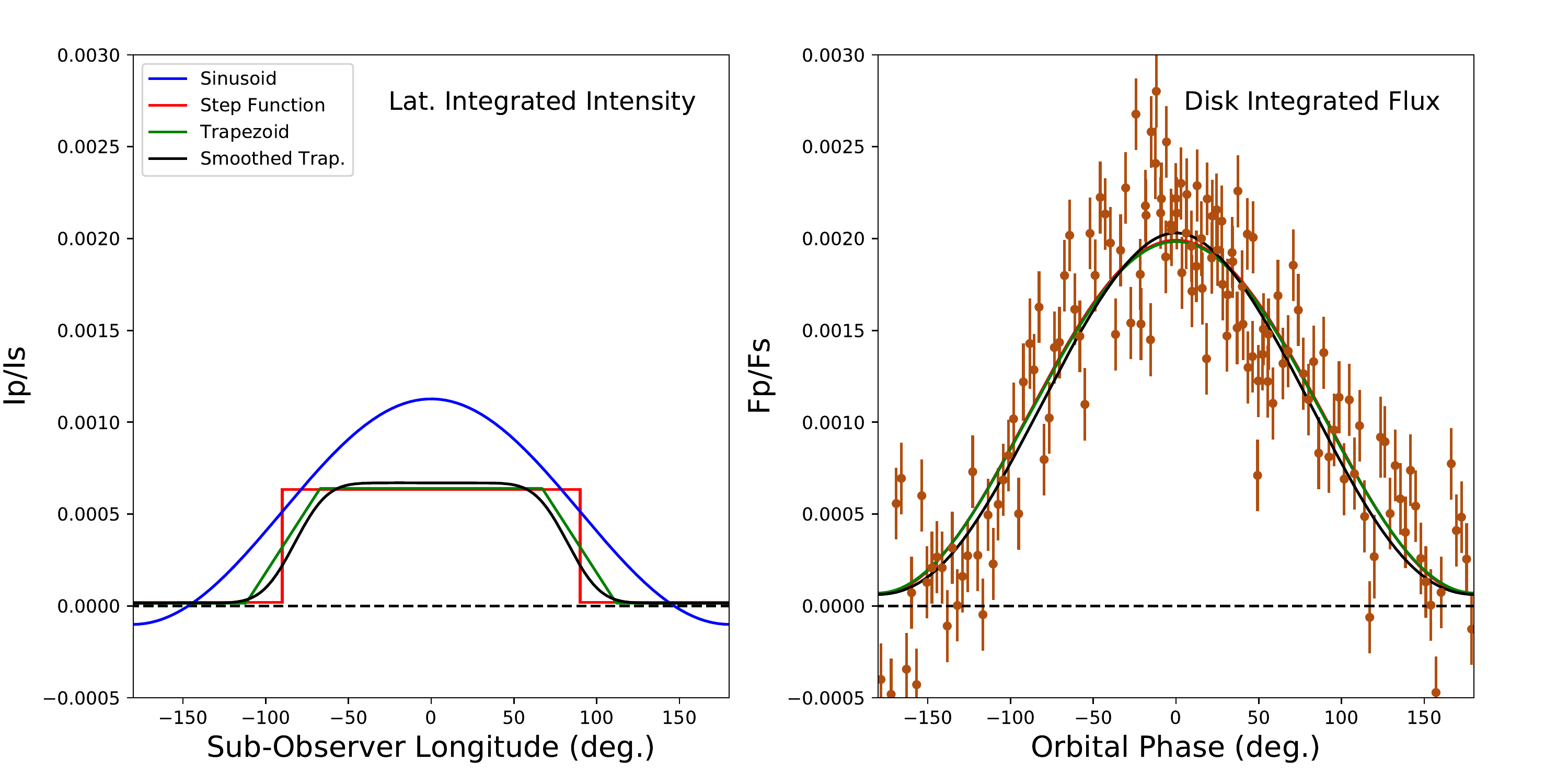}
\vskip -0.0in
\caption{How various latitudinally integrated planetary intensity maps (left panel) integrate to full disk phase curve observations (right panel). Note that the first three integrated models (sinusoidal, step, and trapezoid) in the right panel precisely overlay each other. The data points in the right panel are our \three observations of KELT-1b, with the stellar contributions to the phase curve subtracted away (c.f. Figure \ref{detrend}). For illustration purposes, we have also set the phase offset of the \three observations to $0^\circ$. As noted by \cite{cowan2008}, disparate intensity maps can be integrated to show predicted phase variations indistinguishable in disk-integrated observations, making the inversion of phase curve observations a degenerate problem. Importantly, we can conclude that KELT-1b's intensity map is not sinusoidal, since this would require unphysical, negative, emission on the night side (blue line, left panel). This problem with negative sinusoidal intensity maps was recently noticed by \cite{keating2017}, who suggested it arose from systematics and errors in Spitzer phase curve observations and analyses. Instead, we posit that the negative intensities mean that hot Jupiters' intensity maps are necessarily not sinusoidal, and instead posses a sharp intensity transition between their day- and nightsides -- which is probably caused by nightside cloud formation.}
\label{phaseinvert}
\end{figure*}

Limited and varying dayside clouds could also explain the relatively large scatter shown in Figure \ref{offsets}. \cite{armstrong2016} recently showed that the optical phase offset for HAT-P-7b varied by 80 degrees over 4 years of Kepler observations, which they attributed to changes in reflectivity caused by varying cloud coverage on HAT-P-7b's dayside. Though the effect would be less pronounced at Spitzer wavelengths, such cloud variability would also affect the observed thermal emission on the dayside \citep{powell2018}. 

If hot Jupiters generally have changing dayside cloud cover near the level of the \cite{armstrong2016} observations, then the offsets plotted in Figure \ref{offsets} are not sampling equilibrium thermal process in the atmosphere, but rather are providing us with single snapshots of time-varying weather. Since almost all the planets in Figure \ref{offsets} only have one phase curve observation in each band, such a temporal variability could display itself as a large scatter in the observed offset locations.

A direct way to test for weather-driven scatter in Spitzer phase curve offsets would be to observe multiple phase curves of the same planet at the same wavelength. Ideally one would observe at least three independent phase curves to provide strong evidence of variability, because a differing phase offset between just two observations might simply be the result of unfortunate systematic measurement uncertainties. Since the \cite{armstrong2016} observations varied on timescales of tens to hundreds of days, these repeated phase curve observations should also be spaced apart by two weeks or more.

\subsection{Low Nightside Intensities Required by the Inversion of Phase Curves to Intensity Maps}

Before considering the day-to-night temperature differences in more detail, we next consider the process of ``inverting'' our disk-integrated planet-to-star flux ratios to the underlying atmospheric intensity map. Recently, \cite{keating2017} has pointed out that many Spitzer phase curve observations appear to imply unphysical intensity maps, with negative intensity values on the nightside, to match the disk-integrated phase observations. We found that this was also the case in our KELT-1b measurements (blue curve, left panel of Figure \ref{phaseinvert}), and so we examined this problem in more detail. 

Since we only have longitudinal information about KELT-1b's atmospheric intensity, we sought the latitudinally integrated intensity from the brown dwarf as a function of the sub-observer longitude. For the flux from the star KELT-1, we used a 6500\,K BT-Settl model spectrum for the entire stellar surface\footnote{Note that even under the relatively strong ellipsoidal deformation caused by KELT-1b the effective temperature on the stellar photosphere should not vary by more than about 20\,K (Section 3.3)}. For KELT-1b, unfortunately, the inversion of the observed phase curve to an intensity map is considerably more complicated.

The general forward problem, of transforming a known planetary intensity map to a disk-integrated phase curve, is uniquely solved by computing the integral \citep{rybicki1979}
\begin{equation}\label{eq:42110}
f(\psi) = \int_{\psi-\pi/2}^{\psi+\pi/2} \int_{-\pi/2}^{\pi/2} I(\theta,\phi) \cos^2\phi\cos(\theta-\psi) d\phi d\theta,
\end{equation}
where $\psi$ is the sub-observer longitude, $\theta$ is the planetary longitude, and $\phi$ is the planetary latitude. $I(\theta,\phi)$ is the planetary intensity map as a function of longitude and latitude, and $f(\psi)$ is the resulting disk-integrated phase curve. Since phase curve observations cannot measure latitudinal variation in planetary intensity, we may assume that $I(\theta,\phi)$ is constant as a function of latitude and simplify Equation \ref{eq:42110} to 
\begin{equation}\label{eq:42120}
f(\psi) = \frac{\pi}{2}\int_{\psi-\pi/2}^{\psi+\pi/2} I(\theta) \cos(\theta-\psi) d\theta.
\end{equation}

The inverse problem of going from an observed $f(\psi)$ to $I(\theta)$ does not, unfortunately, have a closed analytic solution. The foundational work on the exoplanet phase curve inversion problem is \cite{cowan2008}, who noted this difficulty. \cite{cowan2008} therefore made the simplifying assumption that the planetary intensity was a linear combination of sinusoidal harmonics as a function of longitude. The resulting observed phase curves are then also a linear combination of sinusoidal harmonics, which allowed \cite{cowan2008} to determine analytic transformations between the amplitudes of the phase curve harmonics and the amplitudes of the intensity map harmonics.  

As mentioned, \cite{keating2017} showed that more Spitzer phase curve observations transform to intensity maps which go negative on the nightside, using the \cite{cowan2008} sinusoidal results. This is also the case for our observations of KELT-1b, which would necessitate a minimum ``temperature'' of roughly -1400\,K assuming a sinusoidal intensity map (blue curve in the left panel of Figure \ref{phaseinvert}). \cite{keating2017} suggested that the prevalence of negative nightside temperatures reflected untreated systematics in Spitzer phase curve observations, and that phase curve analyses should require physical, non-negative, intensity maps as a part of the fitting process.

However, there is an alternate explanation for the negative nightside temperatures measured by \cite{keating2017}: the planetary intensity maps are not sinusoidal as posited by \cite{cowan2008} and as assumed by \cite{keating2017}. Indeed, as shown in the left panel of Figure \ref{phaseinvert}, a variety of non-sinusoidal intensity maps can reproduce our disk-integrated results for KELT-1b without resorting to negative intensities on the nightside. As mentioned above, this basic degeneracy in the inversion problem was noted by \cite{cowan2008}.

As one example, consider a toy-model planetary atmosphere whose dayside is a single hot temperature, and whose nightside is a single cold temperature. For this ``step-function'' atmosphere, if we call the dayside intensity $I_D$ and the nightside intensity $I_N$, we may then use Equation \ref{eq:42120} to determine that the disk-integrated phase curve to be
\begin{equation}\label{eq:42130}
f(\psi) = \frac{\pi}{2}(I_D+I_N)+ \frac{\pi}{2}(I_D-I_N)\cos(\psi).
\end{equation}
A planetary atmosphere with a step-function intensity map would thus show a sinusoidal phase curve observationally indistinguishable from an atmosphere with a sinusoidal intensity map.

Importantly, such a step-function atmosphere would be able to match all the observed Spitzer phase curves without resorting to negative intensities.\footnote{Except for the few cases where the disk-integrated phase curve itself goes to negative flux, such as HAT-P-7b \citep{wong2016}.} Though a perfect step-function intensity map is difficult to imagine existing in reality, Figure \ref{phaseinvert} also shows how more plausible, trapezoidal, intensity functions can replicate the disk-integrated observations without going to negative intensities. In all cases, the necessary requirement to do so is a relatively sharp transition to a nearly constant and lower nightside intensity level such that disk-integrated nightside observations cannot ``see'' the hot dayside atmosphere.

Therefore, the negative nightside temperatures noticed by \cite{keating2017} require both that the underlying atmospheric intensity maps for hot Jupiters are non-sinusoidal, and that there is a sharp day-night intensity transition to a nearly constant nightside. This sort of transition could be caused by a combination of short timescales for atmospheric re-radiation and atmospheric drag \citep{perezbecker2013,komacek2016,komacek2017}, but our KELT-1b observations make this unlikely. If KELT-1b's day-to-night intensity transition were driven solely by thermal processes and the balance between its thermal timescales, we would expect it to show a gradual intensity change as the dayside heat is moved  from day to night. Observationally, however, we see a low nightside flux that requires a sharp intensity transition -- suggestive of non-thermal processes.

\begin{figure}[b]
\vskip 0.00in
\includegraphics[width=1.05\linewidth,clip]{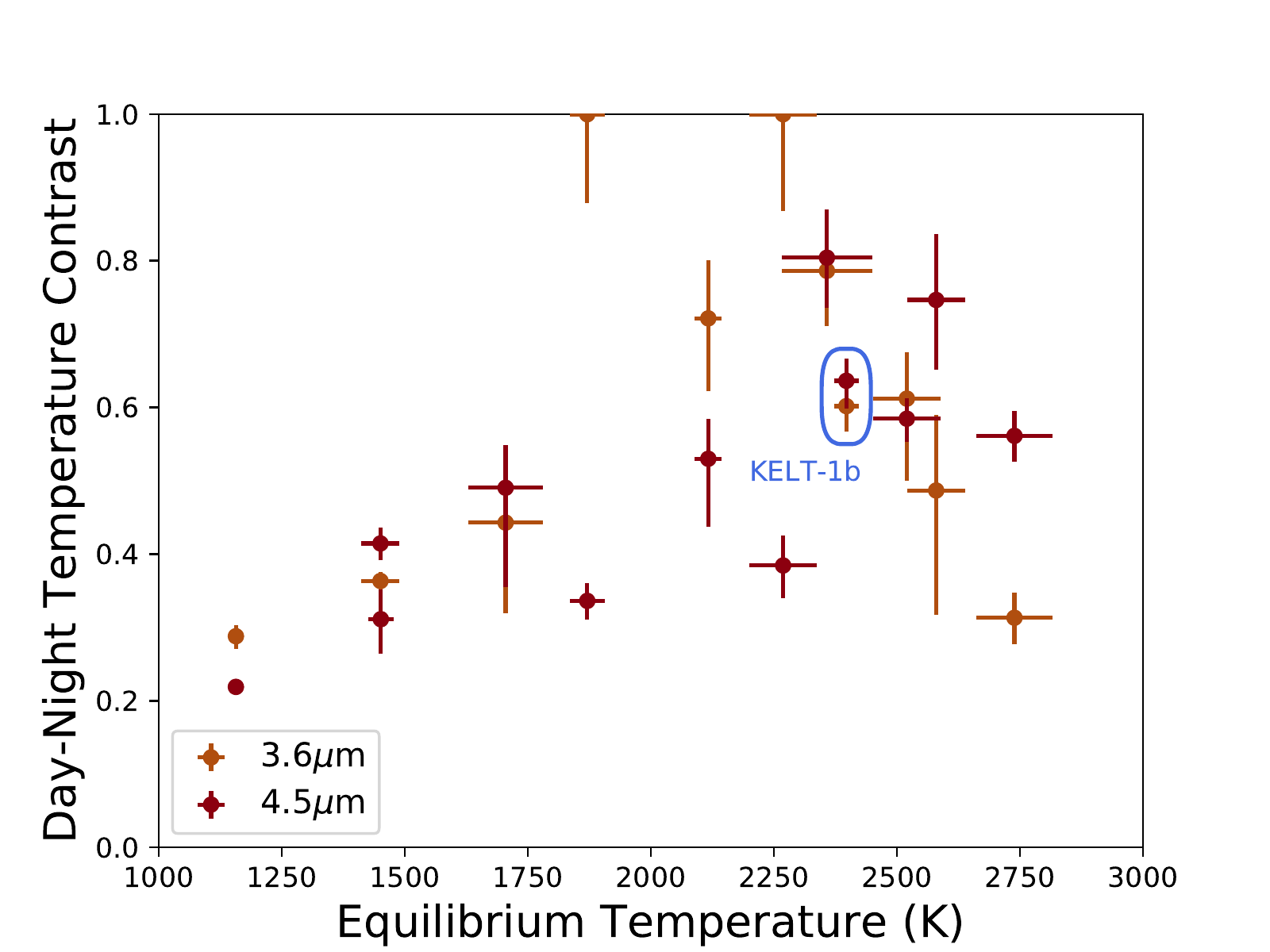}
\vskip -0.0in
\caption{The day-night temperature contrast ($[T_{day}-T_{night}]/T_{day}$) we measure for KELT-1b is approximately 0.62 in both Spitzer bands (the points highlighted by the light blue rectangle), which is in line with the general trend that hot Jupiters with higher equilibrium temperatures show higher contrasts \citep{showman2002,perezbecker2013}. We measure KELT-1b's nightside brightness temperature to be 1173\,K at \three and 1053\,K at \fouralt, which is higher than the brown dwarf's predicted interior luminosity of $\sim900$\,K \citep{saumon2008}, and indicates that the contrast we measured is not being significantly suppressed by emission of internal heat. Compare to Figure \ref{daynight}.} 
\label{contrasts}
\end{figure}

The formation of a nightside cloud deck can account for both of these intensity map requirements. Cloud formation near planetary dusk and dispersal around planetary dawn would act to suddenly decrease the pressure level of the planetary photosphere, moving the observable portion of the atmosphere to lower temperatures \citep{ackerman2001}. A uniform nightside cloud layer would then serve to keep the planetary intensity low and nearly constant across the anti-stellar hemisphere \citep{powell2018}. 

\subsection{Trends in \three and \four Day- and Nightside Brightness Temperatures}

Our observations of KELT-1b show that in both Spitzer channels the brown dwarf displays a relatively large phase amplitude of roughly 1950\,ppm. As listed in Table 2, the minimum fluxes are close to zero -- though due to the offset in both phase curves the flux minima occur after midnight at the planetary ``witching hour'' -- and disk-integrated nightside fluxes are small but significantly non-zero. This gives KELT-1b a dayside brightness temperature of $2988\pm60$\,K and a nightside brightness temperature of $1173_{-130}^{+175}$\,K at \threealt, and a dayside brightness temperature of $2902\pm74$\,K and a nightside brightness temperature of $1053_{-161}^{+230}$\,K at \fouralt. We calculated both sets of brightness temperatures using a 6500\,K BT-Settl \citep{allard2012} model spectrum for the star KELT-1.

\begin{figure*}[t]
\vskip 0.00in
\includegraphics[width=1.05\linewidth,clip]{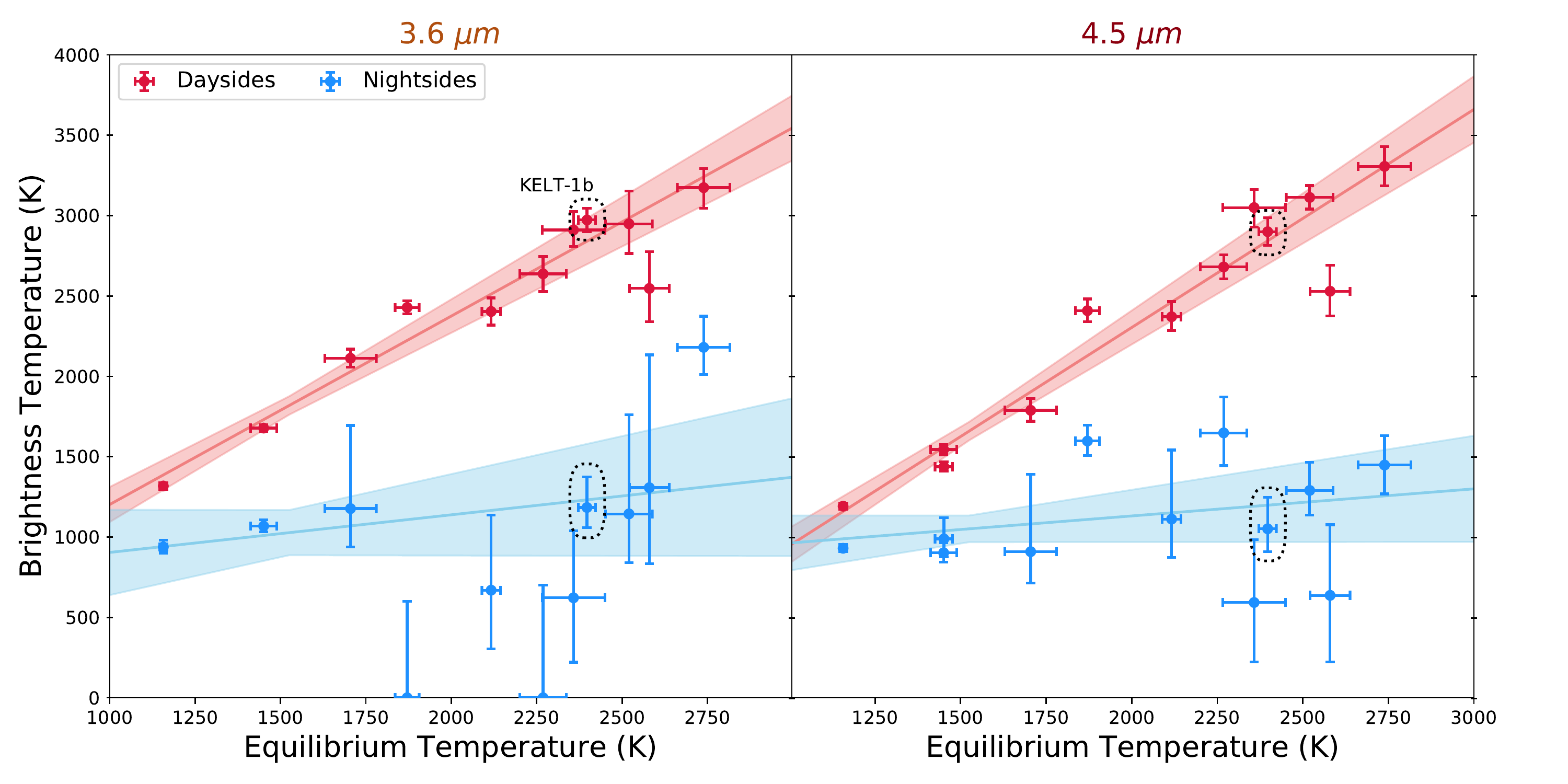}
\vskip -0.0in
\caption{The dayside (red) and nightside (blue) brightness temperatures of hot Jupiters with \three (left) and \four (right) phase curve observations show two trends as a function of planetary equilibrium temperature. Both sets of dayside temperatures increase linearly, with approximately the same slope of 1.25 (Equation \ref{eq:42220}), while the planetary nightside temperatures appear approximately constant at 1000\,K (Equation \ref{eq:42230}). This latter trend is another piece of evidence for nightside clouds, which are presumably ``clamping'' the observed nightside emission temperatures to the temperature of the cloud deck. These trends also provide a new way to view the day-night contrast trend shown in Figure 8. The equilibrium temperatures on the x-axes are the zero-albedo, complete heat redistribution, planetary effective temperatures due to stellar irradiation alone.}
\label{daynight}
\end{figure*}

These temperatures give KELT-1b day-night temperature contrasts of $0.607_{-0.025}^{+0.032}$ at \three and $0.637_{-0.032}^{+0.042}$ at \fouralt. These are consistent with the trend that day-night temperature contrasts for hot Jupiters increase with increasing planetary equilibrium temperature (Figure \ref{contrasts}), which was predicted by \cite{showman2002} and first noted observationally by \cite{perezbecker2013}. Note that the nightside temperatures we measure for KELT-1b are higher than the temperature we would expect from the brown dwarf's interior luminosity, which should give an unirradiated effective temperature of $\sim850$\,K at the \cite{siverd2012} mass and age \citep{saumon2008}. The nightside is therefore emitting somewhere between two to four times more energy than KELT-1b's expected internal luminosity, which indicates that its temperature is being determined by the dayside irradiation.

In addition to the day-night temperature contrasts shown in Figure \ref{contrasts}, we also examined the underlying day- and nightside brightness temperatures for all the hot Jupiters with \three and \four Spitzer phase curves, which are listed at the beginning of Section 4. To calculate the brightness temperatures for the planets, we used BT-Settl model spectra \citep{allard2012} at the corresponding host star effective temperature to estimate the stellar flux, and then the planet-to-star flux ratio in the middle of eclipse and transit to determine the dayside and nightside temperatures, respectively. Note that although many of the planets have nightside flux measurements that are only mildly significant, it is possible to measure the resulting nightside brightness temperature more robustly. We explain this in more detail in Appendix A. We compared these brightness temperatures to the zero-albedo, complete redistribution, blackbody planetary equilibrium temperature -- a proxy for the amount of incoming energy from the host star. 

As shown in Figure \ref{daynight}, we find that in both Spitzer bands the dayside brightness temperatures display a remarkable linear correlation with planetary equilibrium temperature. Specifically, a linear fit to both sets of data of the form $T_{day} = m_{day}(T_{eq}-\bar{T}_{eq}) + b_{day}$ yields
\begin{eqnarray}\label{eq:42210}
m_{day,3.6} &=& 1.13\pm0.08 \quad b_{day,3.6} = 1810\pm44\,\mathrm{K}  \\ \nonumber
m_{day,4.5} &=& 1.33\pm0.11 \quad b_{day,4.5} = 1660\pm46\,\mathrm{K}, 
\end{eqnarray}
where $\bar{T}_{eq}=1524$\,K is the error-weighted mean equilibrium temperature.

Interestingly, a similar linear fit to the nightside brightness temperatures shows that they are approximately constant with equilibrium temperature, as
\begin{eqnarray}\label{eq:42220}
m_{night,3.6} &=& 0.22\pm0.27 \quad b_{night,3.6} = 919\pm153\,\mathrm{K}  \\ \nonumber
m_{night,4.5} &=& 0.12\pm0.15 \quad b_{night,4.5} = 1032\pm76\,\mathrm{K}. 
\end{eqnarray}
That is, the nightsides of all the hot Jupiters are at a roughly constant temperature of 1000\,K at \three and \fouralt.

It is tempting to also conclude that the linear increase in dayside temperatures indicates that all of the hot Jupiters must have similar heat recirculation efficiencies -- and hence atmosphere dynamics. Indeed, the trend for the dayside brightness temperatures is suggestively close to what one would expect if all the planetary atmospheres had a Bond albedo of zero and no heat redistribution from the day- to nightside. If that were the case, we would see
\begin{equation}\label{eq:42230}
m_{no\_redist} = 1.28 \quad b_{no\_redist} = 1958\,\mathrm{K}.
\end{equation}
This slope is roughly consistent with what we see in Figure \ref{daynight} and in Equation \ref{eq:42210}, but the intercept is offset 150\,K to 300\,K higher than the observations. Increasing either the albedo to $A_B=0.3$ or decreasing the redistribution parameter to $f^\prime=0.5$ (following the convention of \cite{seager2010}) in this calculation can lower the intercept $\sim1750$\,K to match the data, though this also decreases the slope to $m\approx1.16$. This would be roughly consistent with both dayside measurements.

However, these data are measuring brightness temperatures at two specific wavelengths, while the usual arguments about recirculation efficiency and Bond albedo pertain to spectrum-wide effective temperature. This has been well established in Spitzer observations of field brown dwarfs, and as an illustration, the dayside of KELT-1b emits just $4\%$ of its total flux at \three, and $2\%$ at \four \citep{beatty2017}. These Spitzer observations thus only sample a fraction of the total dayside emission, and the majority of the thermal emission near 1\,\um\ (and hence the dayside effective temperature) may be modulated by the presence of dayside clouds or other processes unseen at these wavelengths \citep[e.g.,][]{ackerman2001}.

Indeed, if we were to assume that the \three brightness temperatures in Figure \ref{daynight} are the actual effective temperatures of the planetary daysides and nightsides, then on average these planets are emitting 1.2 times more energy than they receive from their host stars. Even at low equilibrium temperatures this level of additional emission would require an internal effective temperature of $\sim800$\,K, which is much hotter than expected for Jupiter-mass objects at these ages \citep{saumon2008}. Thus, the brightness temperatures we see in Spitzer observations must be only roughly related to the actual planetary effective temperatures. This, in turn, makes drawing specific conclusions about the heat redistribution efficiency and Bond albedos of these planets problematic if they are based solely on Spitzer observations, and indicates the need for more accurate measurements of the bolometric luminosities of hot Jupiters \citep{cowan2011b}.

In addition to these brightness versus effective temperature considerations, it is also extremely difficult to see how planetary atmospheric dynamics and timescales would remain constant over this wide range of temperatures to cause all the hot Jupiters to have the same heat recirculation efficiency \citep[e.g.][]{komacek2016,komacek2017}.

These brightness temperature correlations provide a new way to view the well-known trend \citep{showman2002,perezbecker2013} towards higher day-night temperature contrasts at higher equilibrium temperatures (Figure \ref{contrasts}). Namely, that this trend is not primarily the result of changing heat redistribution or radiative and advective timescales in the planetary atmospheres \citep[e.g.,][]{perezbecker2013}, but rather arises because the nightsides of all the hot Jupiters remain at a constant $\sim1000$\,K, while their daysides simply become hotter under increasing stellar irradiation. In principle, both of these trends could be explained by a precise balancing of the atmospheric dynamics in these planets such that increasing stellar irradiation leads to steadily increasing dayside temperatures and constant nightside temperatures, but this seems too coincidental over this wide a range in parameter space.

Instead, the fact that all the observed planetary nightsides are $\sim1000$\,K probably indicates that the thermal emission we see is being set by the formation of a nightside cloud deck at this temperature. In this scenario, the nightside photospheric temperature and pressure probed by \three and \four observations is clamped to the level of the nightside clouds, which must be optically thick at these wavelengths. Based on the temperature of the nightsides, these clouds seem likely to be composed of MnS or Na$_2$S \citep{parmentier2016, powell2018} and have grain sizes $\gtrapprox5$\,\um\ to be optically thick in \three and \four observations. It is also possible that the nightside clouds are composed of Si and are extremely thick: in this scenario we would be seeing just the cloud tops at a lower temperature than the Si condensation point.

\subsection{KELT-1b's Phase Curve Trajectory on a Color-Magnitude Diagram}

\begin{figure}[t]
\vskip 0.00in
\includegraphics[width=1.05\linewidth,clip]{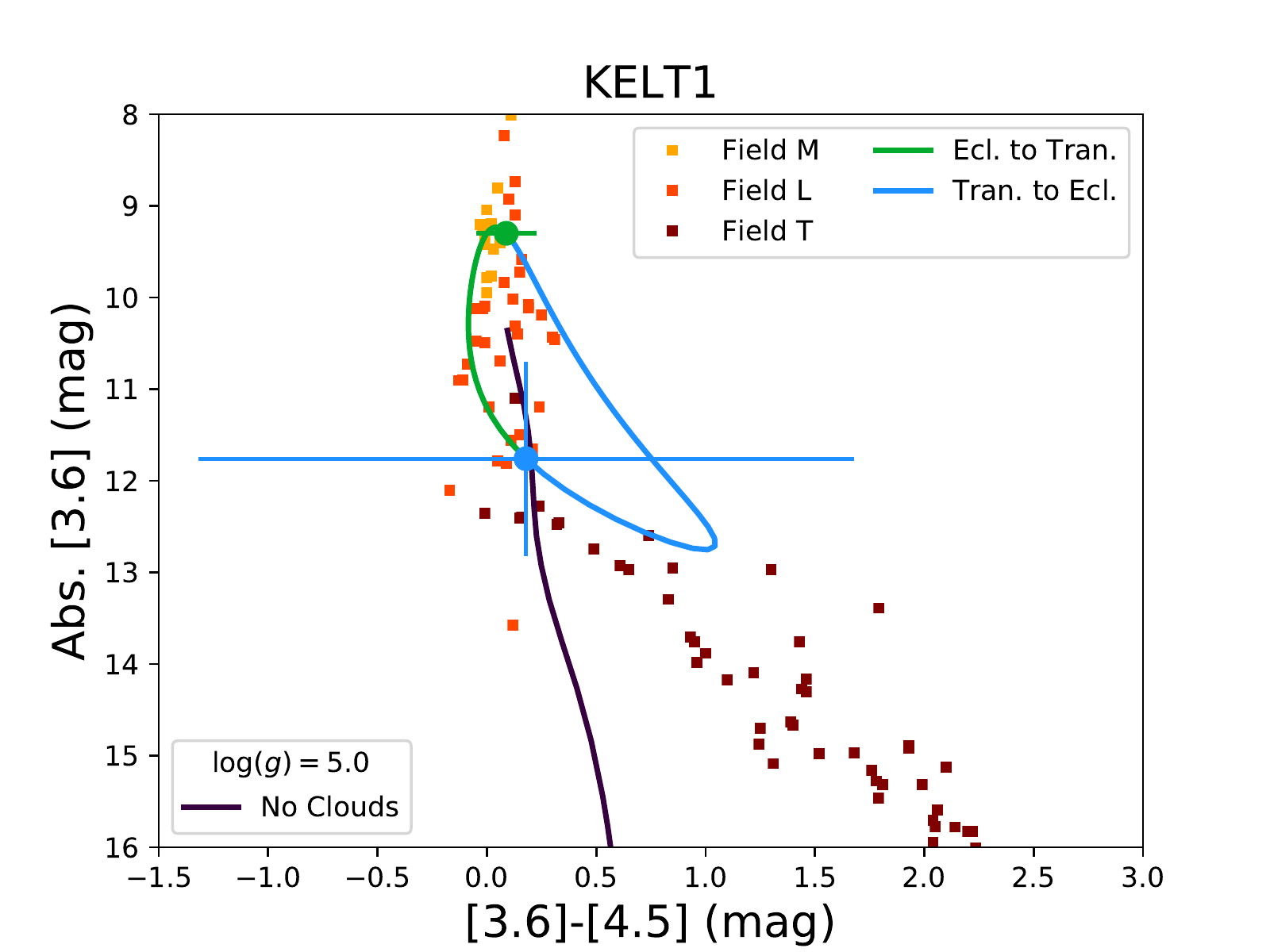}
\vskip -0.0in
\caption{Using our phase curve observations and the Gaia DR2 parallax to KELT-1, we can reconstruct the trajectory KELT-1b's atmosphere takes on a Spitzer CMD over the course of a complete orbit. The large circular points show the location of KELT-1b's dayside (green) and nightside (dark blue), while the background square points are field brown dwarfs \citep{dupuy2012}. In this plot a parcel of gas in KELT-1b's atmosphere travels counter-clockwise around the loop-the-loop. The fact that KELT-1b's nightside follows the brown dwarf sequence below the L-T transition indicates that clouds must be forming on the nightside, since there is not enough time for hot CO-laden gas from the dayside to convert to being \methane\ dominated \citep{cooper2006}..}
\label{kelt1cmd}
\end{figure}

Finally, it is interesting to use the recent Gaia DR2 parallax for KELT-1, and compare the path that KELT-1b's atmosphere takes on a color-magnitude diagram (CMD) over the course of an entire orbit to the field brown dwarf sequence and to brown dwarf atmosphere models. While we typically think of plotting just the day- or nightsides of hot Jupiters on a CMD, phase curve observations give us the unique opportunity plot a complete ``trajectory'' of the atmosphere from day to night, and back again. Thus, in Figure \ref{kelt1cmd} KELT-1b's atmosphere is not just a single (or pair) of points, but rather is described by a ``loop-the-loop'' on the CMD.

When comparing the atmosphere of KELT-1b and hot Jupiters to field brown dwarfs, it is important to keep in mind two key differences. First, the strong external irradiation received by KELT-1b and the planets forces their dayside temperature-pressure (TP) profiles to be hotter than for field brown dwarfs. This also changes the energy balance of the radiative portion of these atmospheres by making the primary energy source be at the top of the atmosphere, rather than at the interior as with a field brown dwarf. Second, KELT-1b and hot Jupiters should lack \methane\ in their atmospheres, even when their nightsides cool below the CO--to--\methane\ interconversion temperature. This is because the  CO--to--\methane\ interconversion timescale in a typical hot Jupiter's atmosphere is more than order of magnitude longer than the nightside crossing time \citep{cooper2006}, ensuring that the hot, CO-laden, atmosphere from the dayside cannot chemically convert to being \methane-dominated.

\begin{figure*}[t]
\vskip 0.00in
\includegraphics[width=1.05\linewidth,clip]{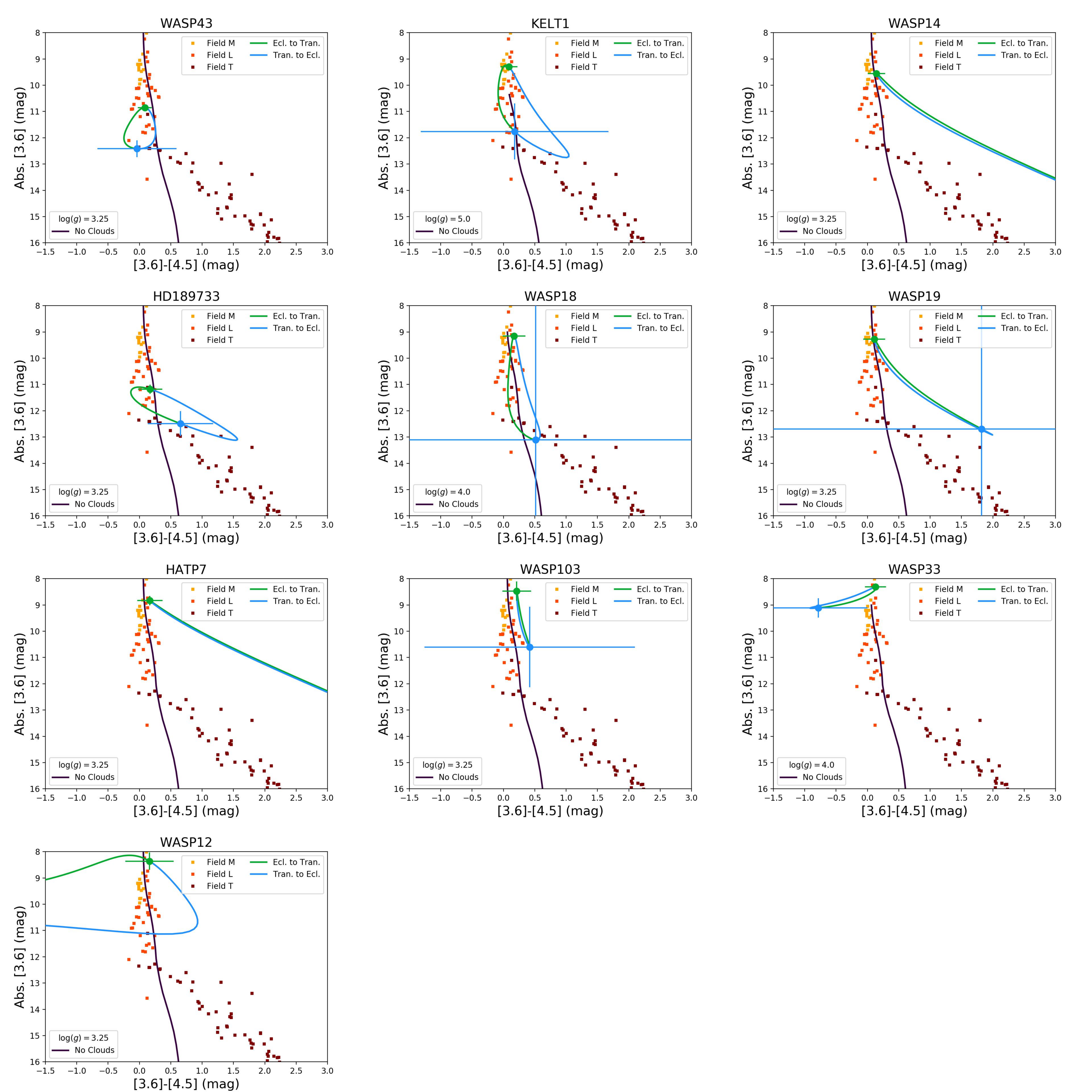}
\vskip -0.0in
\caption{The CMD trajectories for KELT-1b and the other hot Jupiters we compare to in this Section, sorted by signal-to-noise ratio. Much of the variation in the bottom rows is probably due to measurement uncertainty in the planetary phase curves (see the end of Section 4.4 for some of the known issues). That being said, the highest signal-to-noise objects seem to follow the field sequence as they cool, and not the ``No Clouds'' model \citep{saumon2012}. Since there is not enough time for \methane\ to from on the planetary nightsides \citep{cooper2006}, the red nightside colors and trajectories that follow the brown dwarf sequence below the L-T transition are suggestive of cloud formation on these objects' nightsides.}
\label{combinedcmd}
\end{figure*}

That being said, spectroscopic observations of KELT-1b's dayside emission show that it appears indistinguishable from a 3200\,K field ultra cool dwarf spectrum \citep{beatty2017}, and our Spitzer measurements of KELT-1b's dayside place it precisely along the field brown dwarf sequence on a [3.6] vs. [3.6]-[4.5] CMD. As pointed out in \cite{beatty2017}, this spectral similarity indicates that the high surface gravity of KELT-1b is playing the dominant role in setting its dayside TP profile, rather than the irradiation it receives -- other than the fact that the irradiation makes the dayside hotter. Additionally, for half of our phase curve observations -- on KELT-1b's non-irradiated nightside -- we should see a TP profile with boundary conditions similar to a field object. Both these facts make considering the global atmosphere of KELT-1b in the context of field brown dwarfs more of an apples-to-apples comparison.

As shown in Figure \ref{kelt1cmd}, KELT-1b's atmosphere follows a CMD trajectory with two stages. Note that in these CMD trajectory plots a parcel of gas in the atmosphere travels counter-clockwise around the loop-the-loop. Thus, as KELT-1b's atmosphere cools from the hottest part of the day (near eclipse) through dusk, midnight (transit), and towards dawn, Figure \ref{kelt1cmd} shows that it closely follows the field brown dwarf sequence \citep[data from][]{dupuy2012}. After reaching its coolest point shortly before dawn, the atmosphere then rapidly reheats along a nearly isothermal track. 

Interestingly, KELT-1b's nightside CMD trajectory continues along the brown dwarf sequence below the field L-T transition, though given the large uncertainties on the nightside colors shown in Figure \ref{kelt1cmd} the precise trajectory of KELT-1b's atmosphere should be treated as only a general trend.  In field objects this shift in colors is caused by a combination of the switch-over from being CO- to \methane-dominated \citep{burrows1999}, and by cloud formation \citep{ackerman2001}. But as mentioned previously, the CO--to--\methane\ interconversion timescale in KELT1-b's atmosphere is too long for \methane\ to form on the nightside \citep{cooper2006}. This leaves us with clouds as the likely explanation for KELT-1b's nightside CMD trajectory.

To investigate this in more detail, in Figure \ref{kelt1cmd} we have also plotted field brown dwarf atmosphere models from \cite{saumon2012}. As evident in Figure \ref{kelt1cmd}, the ``no clouds'' model does not replicate the observed evolution of KELT-1b's nightside colors, while the field objects with clouds do.

The reddish nightside that we see on KELT-1b is therefore also evidence that clouds are forming and strongly influencing the nightside thermal emission, though given the large uncertainties on KELT-1b's nightside colors, this conclusion from a single object should be treated with caution. We therefore also constructed similar CMD trajectory plots for the hot Jupiters that we have been comparing to in the rest of this Section\footnote{Excepting HD 149026b, since \cite{zhang2018} cautioned against their result at \three}. Individually all of these objects have poorly constrained nightside colors, but we wished to investigate the ensemble properties of their nightside CMD trajectories.

As can be seen in Figure \ref{combinedcmd}, there is a variety of trajectories taken by the hot Jupiters -- not all of which track the field brown dwarf sequence or necessarily make physical sense. The three prominent examples of this are HAT-P-7b, WASP-12b, and WASP-33b. In the case of HAT-P-7b, \cite{wong2016}'s phase curves go negative on the nightside, which presumably reflects unresolved systematics in the dataset. For WASP-12b, \cite{cowan2012} had considerable difficulty modeling the expected ellipsoidal deformation of the planet itself, which influenced their results. Finally, WASP-33b \citep{zhang2018} has a host star which shows noticeable short-period pulsations in the optical, and it is possible that low amplitude pulsations in the Spitzer bandpasses may be complicating measurement of its phase curve.

The six objects in the top two rows of Figure \ref{combinedcmd} give us the highest precision CMD trajectories. One object of particular interest is HD 189733b, which parallels the brown dwarf sequence below the L-T transition and therefore appears to have clouds. This general color shift was noted by \cite{knutson2012} (though not conceptualized this way), who suggested that it was evidence for \methane\ on the nightside due to disequilibrium chemistry. However, \cite{birkby2013} later saw no \methane\ in HD 189733b's emission spectrum. If HD 189733b instead has a red nightside due to clouds, this resolves the discrepancy between the \cite{knutson2012} and \cite{birkby2013} observations.

It is also interesting to note that both WASP-14b and WASP-19b show CMD trajectories above the field brown dwarf sequence. In principle, this could be an indication of lower vertical diffusion rates ($\mathrm{k}_{\mathrm{zz}}$) in their atmospheres compared to the field brown dwarfs \citep{saumon2008}. However, the uncertainties on both trajectories are such that any specific conclusions about these planets' $\mathrm{k}_{\mathrm{zz}}$ values are marginal, at best.

Though the nightside colors of any one of these objects are consistent with both the field sequence and the ``No Clouds'' model, it is important to note that all have either a nominal color of near zero (WASP-43b, WASP-18b) or a nominal color of red (KELT-1b, WASP-14b, HD 189733b, WASP-19b). If we were to assume that the observed nightside colors were purely a result of the measurement noise, and therefore had an even chance of appearing red or blue, then via the binomial distribution these observational results have only a 1.5\% chance of occurring, which implies the ensemble is showing us red nightsides at $2.5\,\sigma$. This significance declines to just $1.9\,\sigma$ if we include all 10 objects, but increases back to $2.3\,\sigma$ is we leave out the hard-to-fit WASP-12b results.

The ensemble CMD trajectories therefore suggest (at 2 to 2.5\,$\sigma$) that all of these objects have red nightside colors and trajectories that follow the brown dwarf sequence below the L-T transition, which is another signature of cloud formation on these objects' nightsides.

\section{Summary and Conclusions}

We used the Spitzer spacecraft to observe two orbital phase curves of the transiting brown dwarf KELT-1b, at \three and \fouralt. Along with previous Spitzer eclipse observations from \cite{beatty2014}, we used BLISS mapping to fit both bands simultaneously using a single set of underlying physical parameters. We strongly detected orbital phase variation from KELT-1b, at a signal-to-noise ratio of approximately 24 in both bands, and measure a significantly non-zero nightside flux (Table 2). 

The resulting day-night temperature contrasts that we measure for KELT-1b are $0.607_{-0.025}^{+0.032}$ at \three and $0.637_{-0.032}^{+0.042}$ at \fouralt, which are in keeping with other observations of hot Jupiters (Figure \ref{contrasts}). We see clear Eastward offsets in the locations of KELT-1b's dayside hotspots, which are above the recent predictions from \cite{zhang2018}, but still in line with the ensemble of other \three and \four offset measurements (Figure \ref{offsets}). We also do not detect any significant short-term variability in KELT-1b's emission, which might underlie the general phase variation itself. 

As we discuss in Section 4, a comparison of KELT-1b to other \three and \four phase curves of hot Jupiters indicates that the observed thermal emission from KELT-1b and these planets is being strongly modulated by the presence of nightside clouds. We describe four lines of evidence for this conclusion:

\begin{enumerate}
\item The available set of \three and \four phase offset measurements -- including KELT-1b -- are consistent with the observed planets having a constant phase offset of 14 deg. for all planetary equilibrium temperatures, though with a high scatter. This conflicts with cloudless GCM predictions that cooler planets should show large ($\sim70$\,deg.) offsets \citep{zhang2018}.
\item The low disk-integrated nightside fluxes measured for KELT-1b and other hot Jupiters require that the underlying latitudinally-average atmospheric intensity map show relatively sharp transition from a hot dayside to a nearly constant and cooler nightside, such that disk-integrated nightside observations cannot ``see'' the hot dayside atmosphere. This requirement solves the problem pointed out by \cite{keating2017} that many hot Jupiters appear to have negative nightside intensities -- if one assumes a sinusoidal intensity map.
\item The day- and nightside brightness temperatures for all the planets at \three and \four show two remarkable trends (Figure \ref{daynight}). First, the dayside brightness temperatures show a clear linear trend as a function of planetary equilibrium temperature. Second, the nightside brightness temperatures in both bands are consistent with all the planets having constant, $\sim1000$\,K, nightsides. This provides a new way to view the well-known trend \citep{showman2002,perezbecker2013} towards higher day-night temperature contrasts at higher equilibrium temperatures (Figure \ref{contrasts}). Namely, that this trend is not primarily the result of changing heat redistribution or radiative and advective timescales in the planetary atmospheres \citep[e.g.,][]{perezbecker2013}, but rather arises because the daysides of hot Jupiters simply become hotter under increasing stellar irradiation, while the nightsides of all the hot Jupiters possess clouds that show nearly uniform $\sim1000$\,K emission. Note that nearly simultaneously with the submission of this manuscript \cite{keating2018} also described these day- and nightside temperature correlations and also suggested that they are evidence for nightside clouds.
\item Using Gaia DR2 parallaxes for KELT-1b and the other planets we can trace the phase evolution of their atmospheres on a color-magnitude diagram (Figure \ref{combinedcmd}). These trajectory plots suggest that most of the planets have nightside colors that are explainable by the presence of clouds \citep{saumon2008}, and not as easily explained by the presence of atmospheric \methane\ \citep{cooper2006}.
\end{enumerate}

In principle, it is possible to explain most of these observational results using cloudless atmosphere models, by specifically adjusting the balance between the atmospheric thermal timescales \citep{komacek2016}. However, the agreement of all these trends over such a wide range of parameter space would require a physical balancing act too coincidental to be easily believable, so we therefore prefer the theory that this is the result of nightside clouds.

As mentioned in the Introduction, modelers of hot Jupiter atmospheres have long expected that clouds should play a noticeable role in setting the thermal emission properties of these planets. However, in the past we have lacked observations of sufficient quantity and precision to justify the Herculean task of constructing GCMs that include the effects of clouds and cloud formation. The arguments above indicate, however, that we have reached a point in our observations of hot Jupiters that cloudless GCM predictions can no longer accurately capture the trends that we see in the data, as particularly evidenced by the apparently constant phase offsets (Figure \ref{offsets}) and nightside temperatures of all the hot Jupiters (Figure \ref{daynight}).

The wide range of possible cloud properties will make modeling their effect on planetary emission a difficult task without detailed observational results. Perhaps the most direct line of attack for this problem would be to observe the nightside emission spectra themselves. For hot Jupiters, though the signature of the nightside emission is present in spectroscopic transit observations, the typical transmission spectroscopy signal should be much higher than the nightside emission signal in nearly all cases. KELT-1b is a notable exception to this since as a brown dwarf its transmission spectroscopy signal will be close to zero. One way to directly observe nightside emission spectra would therefore be to target transiting hot brown dwarfs. More generally, for the hot Jupiters one will need spectroscopic phase curve observations to identify the planetary emission levels in transit. Though time intensive, spectroscopic phase curve observations do have the added benefit of allowing us to directly see how clouds change the spectra of individual planets as they switch from clear to overcast skies, which provide invaluable input into modeling the cloud formation processes on these planets.

Another potential method to observe the cloud dynamics on hot Jupiters would be to conduct repeat observations of the same targets. As noted in Section 4.1, nearly all of the planets that have Spitzer phase curve observations have only a single measurement in each \textsc{irac} channel. Repeated broadband phase curve observations could therefore allow us to determine the time-variability in hot Jupiter atmospheres, which presumably would be driven by changing clouds (i.e., weather). As described in Section 4.1, this would allow us to determine if the lack of a trend in the measured phase offsets seen in Figure \ref{offsets} is actually the result of weather-driven scatter in the observations. An alternate way to gauge the presence of weather on hot Jupiters would be to take repeated secondary eclipse spectra. The toy model proposed by \cite{armstrong2016} to explain the variability in HAT-P-7b's optical phase curve requires substantial cloud-driven modulation of the dayside thermal emission, which might be detectable in precise eclipse spectra collected at separated times.

Finally, more indirect methods may also allow us to constrain some of the cloud properties in hot Jupiters. \cite{beatty2017a} suggested that determining the point at which atmospheric cold traps become effective could constrain the average cloud particle size on planetary nightsides. Precise eclipse mapping of a planetary dayside \citep[e.g.,][]{rauscher2018} would reveal both the longitudinal and latitudinal surface intensity maps for the dayside, potentially resolving the degeneracies in inverting phase curves to global intensity maps described in Section 4.2.

\ 
\acknowledgements

This work is based on observations made with the Spitzer Space Telescope, which is operated by the Jet Propulsion Laboratory, California Institute of Technology under a contract with NASA. Support for this work was provided by NASA. 

T.G.B. was partially supported by funding from the Center for Exoplanets and Habitable Worlds. The Center for Exoplanets and Habitable Worlds is supported by the Pennsylvania State University, the Eberly College of Science, and the Pennsylvania Space Grant Consortium.

This work has made use of NASA's Astrophysics Data System, the Exoplanet Orbit Database and the Exoplanet Data Explorer at exoplanets.org \citep{exoplanetsorg}, the Extrasolar Planet Encyclopedia at exoplanet.eu \citep{exoplanetseu}, the SIMBAD database operated at CDS, Strasbourg, France \citep{simbad}, and the VizieR catalog access tool, CDS, Strasbourg, France \citep{vizier}. 

\facility{Spitzer (IRAC)}

\software{Astropy \citep{astropy}, BATMAN \citep{kreidberg2015}, emcee \citep{dfm2013}}

\appendix

\section{Spitzer's Brightness Temperature Uncertainty in the Wien Tail}

Many of the nightside brightness temperatures shown in Figure \ref{daynight} are calculated from \three and \four nightside flux measurements that are close to zero and usually only detected with mild significance. Nevertheless, most of the brightness temperatures in Figure \ref{daynight} are clearly non-zero. KELT-1b's nightside is a good example of this: in both channels the nightside fluxes are measured to be approximately $4\,\sigma$ above zero, but KELT-1b's measured nightside brightness temperatures are nearly $10\,\sigma$ above zero.

This occurs because in both Spitzer channels brightness temperatures cooler than approximately 1000\,K are sampling the Wien Tail of the blackbody function. The sharp slope of the Wien Tail then means that small changes in the implied brightness temperature yield large changes in the observed flux, and vice versa.

Mathematically, consider Wein's Law:
\begin{equation}\label{eq:A1}
I = \frac{2 h c^2}{\lambda^5}\,e^{\frac{-hc}{\lambda kT}}.
\end{equation}
For a given observed flux $I$, the brightness temperature in the Wien Tail will then be
\begin{equation}\label{eq:A2}
T_\mathrm{b} = \frac{h c}{k \lambda}\,\left(\ln\left[1+\frac{2 h c^2}{I \lambda^5}\right]\right)^{-1}.
\end{equation}
The uncertainty on $T_b$ is therefore
\begin{equation}\label{eq:A3}
\sigma_{T_b} = \frac{\partial T_b}{\partial I} \sigma_I = \frac{2 h^2 c^3}{\lambda^6}\frac{\sigma_I}{I\left(\frac{2 h c^2}{\lambda^5}+I\right)}\left(\ln\left[1+\frac{2 h c^2}{I \lambda^5}\right]\right)^{-2}.
\end{equation}

Now let us say that we have measured the observed flux with some fraction uncertainty $f$, so that $\sigma_I / I = f$. In Equation \ref{eq:A3} we can substitute this for $\sigma_I$, and then divide by Equation \ref{eq:A2} to arrive at the resulting fractional uncertainty on the brightness temperature,
\begin{equation}\label{eq:A4}
\frac{\sigma_{T_b}}{T_b} = \frac{2 h c^2}{\lambda^5}\frac{f}{\frac{2 h c^2}{\lambda^5}+I}\left(\ln\left[1+\frac{2 h c^2}{I \lambda^5}\right]\right)^{-1}.
\end{equation}
We can then use Equation \ref{eq:A1} to substitute for $I$, which gives us
\begin{equation}\label{eq:A5}
\frac{\sigma_{T_b}}{T_b} = \frac{f}{1+e^{\frac{-hc}{\lambda kT}}}\left(\ln\left[1+e^{\frac{-hc}{\lambda kT}}\right]\right)^{-1}.
\end{equation}
This can be simplified to 
\begin{equation}\label{eq:A6}
\frac{\sigma_{T_b}}{T_b} \approx f\,\frac{\lambda k T}{h c}.
\end{equation}
If we now say that $T=1000$\,K and $\lambda=4$\,\um\ (i.e., the average wavelength of the two Spitzer channels), then Equation \ref{eq:A6} reduces to
\begin{equation}\label{eq:A7}
\frac{\sigma_{T_b}}{T_b} = 0.28\,f.
\end{equation}

Which is to say that if Spitzer measures a nightside flux with a fractional uncertainty $f$ in the Wien Tail, the fractional uncertainty on the associated brightness temperature will be approximately three times smaller. As mentioned previously, we see this in KELT-1b: the nightside fluxes are measured to be approximately $4\,\sigma$ above zero, and the nightside brightness temperatures are $10\,\sigma$ above zero. Thus we can robustly measure the nightside brightness temperatures shown in Figure \ref{daynight} using flux measurements that are only mildly significant.

\end{document}